\documentclass[11pt]{article}

\pdfoutput=1

\usepackage[utf8]{inputenc}
\usepackage[T1]{fontenc}

\usepackage[margin=1.4in]{geometry}

\usepackage{mathpazo}
\usepackage{eulervm}
\usepackage[shortcuts]{extdash}

\usepackage{amsmath}
\usepackage{amssymb}

\usepackage{amsthm}

\usepackage{mathtools}
\usepackage{mathrsfs}
\usepackage{hyperref}
\usepackage{enumitem}
\usepackage{authblk}
\setlist[enumerate]{topsep=0pt,itemsep=0pt}

\usepackage{graphicx}
\usepackage{overpic}
\usepackage{tikz}
\usetikzlibrary{decorations.markings}
\usepackage{subfig}
\usepackage{makecell}

\graphicspath{{Figures/}}

\numberwithin{equation}{section}

\hypersetup{
colorlinks=true,
linkcolor=blue,
citecolor=blue
}

\theoremstyle{definition}
\newtheorem{obs}{Observation}

\theoremstyle{plain}
\newtheorem{example}{Example}
\newtheorem{prop}{Proposition}
\newtheorem{theorem}{Theorem}

\newcommand{\Eqref}[1]{Eq.~\ref{#1}}
\newcommand{\Figref}[1]{Figure~\ref{#1}}
\newcommand{\Propref}[1]{Proposition~\ref{#1}}
\newcommand{\Theoref}[1]{Theorem~\ref{#1}}
\newcommand{\Exref}[1]{Example~\ref{#1}}
\newcommand{\Secref}[1]{Section~\ref{#1}}
\newcommand{\Appref}[1]{Appendix~\ref{#1}}

\newcommand{\dd}{\mathrm{d}}
\newcommand{\disc}[1]{\mathrm{Disc}\,#1}
\newcommand{\plus}{+\!}
\newcommand{\minus}{-\!}

\title{\textsc{Resurgence, a problem of missing exponential corrections in asymptotic expansions}}
\author{Ramon Miravitllas Mas}
\affil{IFAE-BIST, Universitat Autònoma de Barcelona\\
E-08193 Bellaterra (Barcelona), Spain\\
\texttt{rmiravitllas@ifae.es}}

\begin{document} 
\maketitle

\begin{abstract}
\noindent
It is well known that perturbative expansions of path integrals are divergent. These expansions are to be understood as asymptotic expansions, which encode the limiting behaviour of the path integral for positive small coupling. Conventionally, the method of Borel summation assigns a finite answer to the divergent expansion. Still, the Borel sum might not encode the full information of a function, because it misses exponentially small corrections. In the present work, we consider a slight variation of the conventional Borel summation, in which a generalised Borel transform (an inverse Laplace transform) is followed by a directional Laplace transform. These new tools will allow us to give perhaps better answers to typical problems in Borel summation: missing exponential corrections and ambiguities in the Borel summation. In addition, we will define resurgence as a connection between the discontinuity of a function and the coefficients of its asymptotic expansion. From this definition, we will be able to reduce resurgence to the problem of missing exponential corrections in asymptotic expansions and understand, within a unified framework, different approaches to resurgence found in the literature.
\end{abstract}

\section{Introduction}
We consider a general field theory whose action is $S[\phi,g]$, where $\phi$ is a field and $g$ is a coupling that parametrises the strength of the field interaction. The expectation value of an arbitrary functional $\mathcal{O}[\phi]$ is given by the Euclidean path integral
\begin{equation}
\langle \mathcal{O}\rangle(g) = \frac{1}{Z(g)} \int [\dd\phi]\, e^{-S[\phi,g]}\, \mathcal{O}[\phi]\,,
\label{eq_connection_intro_O_path}
\end{equation}
where $Z(g)$ is the partition function of the theory.

We assume that the free theory ($g=0$) reduces to a multivariate normal distribution, so we have a method to compute any desired expectation value of a polynomial in the fields (either Isserlis' theorem or Wick's theorem). Using perturbation theory, we may then compute expectation values in the interacting theory as an expression in powers of $g$:
\begin{equation}
\langle \mathcal{O}\rangle(g) \sim \sum_{n\ge 0} a_n g^{n}\,, \quad g\rightarrow 0^+,
\label{eq_connection_intro_asymptotic}
\end{equation}
which is to be understood as an asymptotic expansion (in the mathematical sense) for $\langle \mathcal{O}\rangle(g)$, rather than a Taylor expansion around $g=0$. In fact, the series diverges for all values of $g\neq 0$. The coefficients $a_n$, given by combinations of expectation values in the free theory, are factorially divergent \cite{Beneke,Dyson,Lipatov}.

Resurgence is concerned with whether the asymptotic expansion in \Eqref{eq_connection_intro_asymptotic} encodes, in some way, the full information about $\langle \mathcal{O}\rangle(g)$ as a function of $g$. Even if we will base the discussion around path integrals, this question clearly makes sense for any arbitrary function, no matter where its asymptotic expansion comes from.

In physics, resurgence  has been gathering attention over the past years. It has applications in quantum mechanics \cite{Justin}, matrix models \cite{Marino2014}, supersymmetric gauge theories \cite{AnicetoRussoSchiappa} and topological string theory \cite{AnicetoSchiappaVonk}. For a very exhaustive list of references, see the introduction in \cite{AnicetoBasarSchiappa}.

Conventionally, the divergent asymptotic expansion in \Eqref{eq_connection_intro_asymptotic} is resummed to a finite answer by a process known as Borel summation. This process consists of two steps. First, we compute the Borel transform of the asymptotic series, given by $\sum a_n\zeta^n/n!$. Second, we \guillemotleft reverse\guillemotright{} the first step by computing the Laplace transform of the Borel transform. This Laplace transform, known as the Borel sum, now converges for $g > 0$ and shares the same asymptotic expansion as the original function.

However, there is no reason to believe that the Borel sum of the asymptotic expansion coincides with the original function. In fact, as much as both of these functions share the same asymptotic expansion, they might still differ by exponentially suppressed corrections that are hidden beyond all terms of this expansion.\footnote{Actually, this is even true for convergent asymptotic expansions. Two functions might share the same convergent asymptotic expansion, but still differ by an exponential correction. The problem arises not because an expansion is divergent, but because it is asymptotic.}

In addition, the large order behaviour of the coefficients $a_n$ produce singularities in the Borel transform. These singularities may give rise to exponentially suppressed ambiguities in the resummation procedure. Thus, we are forced to add exponential corrections to the original asymptotic expansion so as to cancel these ambiguities.

Instead of the asymptotic expansion in \Eqref{eq_connection_intro_asymptotic}, we consider the following extension, known as a transseries, which incorporates potentially missing exponential corrections,
\begin{equation}
\langle \mathcal{O}\rangle(g) \sim \sum_{n\ge 0} a_n g^{n} + i e^{-S_1/g}(-g)^{-\alpha}\sum_{n\ge 0} b_n g^{n} + \dots\,, \quad g\rightarrow 0^+,
\label{eq_connection_intro_transseries}
\end{equation}
where $S_1>0$ and the dots might contain additional exponential corrections, like $e^{-S/g}$, with $S>S_1$.

The coefficients $a_n$ are obtained from quantum fluctuations around the trivial saddle point of the action $S$. That is, a field configuration $\Phi$ for which $\delta S[\Phi,g]/\delta \phi = 0$ and $S[\Phi,g]=0$. The coefficients $b_n$ are computed from non-trivial saddle points, with $S[\Phi,g] = S_1/g \neq 0$.

At first sight, it seems like there is no way that the original asymptotic expansion encodes the full information of the function. Clearly, the coefficients $b_n$ may be completely decoupled from the $a_n$. Nevertheless, in the context of path integrals, there is indeed a connection between the two sets of coefficients. As described in \cite{DunneUnsal}, resurgence is the connection between the large order behaviour of the coefficients $a_n$ and the low order behaviour of the coefficients $b_n$ (and, in fact, this connection also happens between different exponential sectors of the transseries). In this sense, the information in $b_n$ and in the coefficients of other exponential sectors is redundant. In other words, the asymptotic expansion in \Eqref{eq_connection_intro_asymptotic} fully encodes the function.

Closely related, in quantum mechanics, the energy levels of a Hamiltonian can be written as a 1-dimensional Euclidean path integral. For example, the ground energy is given by
\begin{equation}
E(g) = \lim_{T\rightarrow +\infty} -\frac{1}{T} \log\mathrm{Tr}\left( e^{-HT} \right),
\label{eq_connection_intro_E_path}
\end{equation}
where
\begin{equation}
\mathrm{Tr}\left( e^{-HT} \right) = \int_{q(0)=q(T)} [\dd q]\, e^{-S[\phi,g]}
\label{eq_connection_intro_E_path_2}
\end{equation}
and $S[\phi,g]$ is the field version of the Hamiltonian $H$ integrated in the time interval $[0,T]$. As in the case of the field theory, we may compute an asymptotic expansion
\begin{equation}
E(g) \sim \sum_{n\ge 0} a_n g^n\,, \quad g\rightarrow 0^+\,.
\label{eq_connection_intro_E_asymptotic}
\end{equation}
In this scenario, $E(g)$ has a branch cut along $\mathbb{R}^-$ and resurgence is understood as a connection between the leading behaviour, for $g\rightarrow 0^-$, of the discontinuity along the cut $\disc{E}(g)$ and the large order behaviour of the coefficients $a_n$ in \Eqref{eq_connection_intro_E_asymptotic} \cite{Justin,BenderWu,CollinsSoper}.\footnote{Perhaps it is even better to think that the connection is in fact between the coefficients of the asymptotic expansion of $\mathrm{Tr}\left( e^{-HT} \right)$ and the discontinuity of this function. As argued in \cite{Justin}, this connection is then inherited by $E$ through the relation in \Eqref{eq_connection_intro_E_path}.}

At the same time, the asymptotic behaviour of $\disc{E}(g)$ can be computed from the non-trivial saddle points of the action appearing in \Eqref{eq_connection_intro_E_path_2} \cite{Justin,CollinsSoper}. In particular, if $\Phi$ is a saddle point with $S[\Phi,g] = S_1/g$, then
\begin{equation}
\disc{E}(g) \sim 2i\, e^{-S_1/g} (-g)^{-\alpha} \sum_{n\ge 0} b_n g^n\,, \quad g \rightarrow 0^-,
\label{eq_connection_intro_im}
\end{equation}
for some $b_n$ and $\alpha$.

To recapitulate, we have seen that there are two different ways to understand resurgence. In the first case, in a general field theory, we have a connection between the coefficients $a_n$ and $b_n$ in the transseries of \Eqref{eq_connection_intro_transseries}. In the second case, in quantum mechanics, we have a connection between the coefficients $a_n$ in the asymptotic expansion of \Eqref{eq_connection_intro_E_asymptotic} and the coefficients $b_n$ in \Eqref{eq_connection_intro_im}, which encode the asymptotic behaviour of the discontinuity. In the present work, we will argue that both interpretations of resurgence are two sides of the same coin. 
%In essence, the exponential corrections in the transseries of \Eqref{eq_connection_intro_transseries} encode the discontinuity of the function $\langle \mathcal{O} \rangle(g)$.

As a preparation, in \Secref{sec_connection_notation} we will introduce the tools and the notation that we will use thorough this work. The new tools are just extensions of the conventional Borel transform and Borel sum. In this section, we will argue why it makes sense to consider these generalisations.

In \Secref{sec_connection_nevanlinna}, we will discuss sufficient conditions that forbid the existence of exponential corrections hidden beyond the asymptotic expansion of a function. This result is known as Nevanlinna's theorem and it will be central to our discussions and conclusions in the forthcoming sections.

In \Secref{sec_connection_formal}, under the framework of Borel and Laplace transforms, we will formalise the connection previously described between the asymptotic expansion of the ground energy $E(g)$ and its discontinuity (first presented in \cite{BenderWu} and valid when $E(g)$ satisfies a dispersion relation in the $g$ plane). We will call this result a \guillemotleft resurgent connection\guillemotright{}, because it is a connection between the trivial saddle point of the action $S$ and the non-trivial saddle points. To introduce the reader in this topic, we will first review the work of \cite{BenderWu} in \Secref{sec_connection_heuristic}.

Also, in \Secref{sec_connection_formal}, we will show that the discontinuity of a function determines the singularities of its Borel transform. Conversely, singularities in the Borel plane determine the discontinuity of the function.

\Secref{sec_connection_median} contains the main results of this work. There we will exemplify that \guillemotleft median resummed series\guillemotright{} still feature the resurgent connection even if those functions never satisfy a dispersion relation. Then, for median resummed series, we will bring together two apparently unrelated features: the resurgent connection as described in \cite{BenderWu} and the connection between the coefficients $a_n$, $b_n$ in the transseries of \Eqref{eq_connection_intro_transseries}.

%By combining heuristic and formal approaches, we will
%\begin{enumerate}[label=(\arabic*)]
%\item \label{it_connection_main_(1')} Discuss sufficient conditions that forbid the existence of exponential corrections hidden beyond the asymptotic expansion of a function (\Secref{sec_connection_nevanlinna}).
%\item \label{it_connection_main_(1)} Improve on the original discussion from \cite{BenderWu} that guarantees the resurgent connection (\Secref{sec_connection_heuristic}).
%\item \label{it_connection_main_(2)} Show that the discontinuity of a function determines the singularities in its Borel plane. Conversely, singularities in the Borel plane determine the discontinuity of the function (\Secref{sec_connection_formal}).
%\item \label{it_connection_main_(3)} Exemplify that \guillemotleft median resummed series\guillemotright{} still feature the resurgent connection even if those functions never satisfy a dispersion relation (\Secref{sec_connection_median}).
%\item \label{it_connection_main_(4)} For median resummed series, bring together two apparently unrelated features: the present resurgent connection and the connection between the coefficients $a_n$, $b_n$ in the transseries of \Eqref{eq_connection_intro_transseries} (\Secref{sec_connection_median}).
%\end{enumerate}
%Points \ref{it_connection_main_(3)} and \ref{it_connection_main_(4)} are the main results of this work.

\section{Tools and notation}
\label{sec_connection_notation}
%We do not know how to compute the analytic expressions for many relevant quantities in quantum field theories and, at most, we are only able to understand its properties given some limit in the relevant variable of the problem. The behaviour of the function in the limit is encoded in an asymptotic expansion like:

Because the results presented in this work hold for a variety of situations, we will consider a generic complex analytic function $f$ with a power-like asymptotic expansion given by
\begin{equation}
f(z) \sim \sum_{n\ge 0} \frac{a_n}{z^{n+1}}\,, \quad z\rightarrow +\infty\,, \label{eq_connection_f_series}
\end{equation}
where the coefficients $a_n$ are factorially divergent, so the series diverges for all $z$.

When relevant, we will frame the discussion again in the context of physics. In particular, $f$ will be an Euclidean path integral (like \Eqref{eq_connection_intro_O_path} or \Eqref{eq_connection_intro_E_path_2}) and $z=1/g$.\footnote{On some occasions, for the discussion of quark-hadron duality (see \cite{Shifman} for an introduction), $f$ will be a two-point correlator in quantum chromodynamics and $z=q^2$, where $q$ is the (large) momentum going through the correlator. In this case, \Eqref{eq_connection_f_series} has to be understood as the operator product expansion of the correlator and, in all generality, the coefficients $a_n$ can contain logarithms of $z=q^2$.} In perturbation theory, we compute an asymptotic expansion of $f$ in powers of small positive $g$, which corresponds precisely to the expansion in \Eqref{eq_connection_f_series}.

It is standard to define the Borel transform of the asymptotic series in \Eqref{eq_connection_f_series}:
\begin{equation}
B(\zeta) = \sum_{n\ge 0} \frac{a_n\zeta^n}{n!}\,.
\label{eq_connection_conventional_borel}
\end{equation}
Because the $a_n$ are factorially divergent, this function converges in a disc around 0. If we can analytically continue the Borel transform to a strip around $\mathbb{R}^+$, then we may verify that the Laplace transform of $B$ satisfies
\begin{equation}
\int_0^\infty \dd\zeta\, e^{-z\zeta}\, B(\zeta) \sim \sum_{n\ge 0} \frac{a_n}{z^{n+1}}\,, \quad z\rightarrow +\infty\,.
\label{eq_connection_laplace}
\end{equation}
by integrating \Eqref{eq_connection_conventional_borel} term by term. The formalisation of this result is part \ref{it_connection_nevanlinna(b)} of Nevanlinna's theorem below.

Given that $f$ and the above Laplace transform both have the same asymptotic expansion, one hopes the two functions coincide. However, this might not be true in general, because two functions that differ by an exponentially small term, like $e^{-z}$, still share the same power-like asymptotic expansion. Part \ref{it_connection_nevanlinna(a)} of Nevanlinna's theorem gives sufficient conditions to ensure that these exponential corrections are not present and, thus, to ensure that $f$ coincides with the Laplace transform in \Eqref{eq_connection_laplace}.

In the present work, instead of the conventional Borel transform in \Eqref{eq_connection_conventional_borel}, we consider the inverse Laplace transform
\begin{equation}
B(\zeta) = \frac{1}{2\pi i} \int_{\mathcal{C}_a}\! \dd z\, e^{z\zeta}\, f(z)\,,
\label{eq_connection_borel}
\end{equation}
where $\mathcal{C}_a$ is the path $a + iy$, $y\in \mathbb{R}$, with $a$ a constant to the right of all singularities of $f$.

There are two reasons for this redefinition. First, it applies whether $f$ admits the asymptotic expansion in \Eqref{eq_connection_f_series} or not. Second, it clearly reveals that the Borel transform $B$ is related to the singularities of $f$ and, in particular, to its discontinuity.

One can check that this definition coincides with \Eqref{eq_connection_conventional_borel} if the $f(z)$ appearing in \Eqref{eq_connection_borel} is replaced by its power-like asymptotic expansion and each term $1/z^{n+1}$ is integrated with the residue theorem. In this sense, the inverse Laplace transform in \Eqref{eq_connection_borel} is an extension on the initial definition of the Borel transform.

Instead of the conventional Laplace transform in \Eqref{eq_connection_laplace}, we will consider different directions of integration:
\begin{equation}
\int_0^{\infty e^{i\theta}}\! \dd\zeta\, e^{-z\zeta}\, B(\zeta)\,,
\label{eq_connection_directional_laplace}
\end{equation}
parametrised by $\theta$. In the standard discussion on Borel summation, only two paths, slightly above and slightly below the positive real axis, are considered. These two paths can be obtained by taking $\theta$ close to 0. Let us argue why considering arbitrary directions $\theta$ will be convenient.

In order to correctly identify the presence of exponential corrections hidden from an asymptotic expansion, we need to understand what happens in the whole complex plane of $z$. Intuitively, while exponential corrections might be hidden for $z\rightarrow +\infty$, they might become enhanced by taking the limit in an arbitrary direction: $z\rightarrow \infty e^{i\theta}$.

Our first observation is that, if $|B(\zeta)|\le K e^{A|\zeta|}$ (we say $B$ is exponentially bounded), then
\begin{equation}
\int_0^{\infty} |\dd\zeta|\, \left|e^{-z\zeta}\right|\, |B(\zeta)| \le K \int_0^\infty \dd\zeta\, e^{-(\Re(z)-A)\zeta}
\end{equation}
and the last integral converges if and only if $\Re(z)-A>0$. Absolute convergence ensures that the Laplace transform defines an analytic function in $\Re(z)>A$. Now, changing the direction $\theta$ in the Laplace transform, we attain the regions of absolute convergence $\Re\big( z e^{i\theta} \big) > A$. If no singularities are located between the directions $\theta$ and $\theta=0$, then the original Laplace transform yields the same result as the directional Laplace transform in the intersection $\Re(z)>A \cap \Re\big( z e^{i\theta} \big) > A$. Thus, the directional Laplace transform provides analytical continuations to the original Laplace transform. Changing the direction of integration lets us explore the whole complex plane of $z$ and, in fact, some regions of the Riemann surface of the function.

This has two main advantages. First, We will be able to extend the validity of the original asymptotic expansion from $z>0$ to different regions of the Riemann surface. Second, we will understand the imaginary ambiguities that may arise in conventional Borel summation in the following way: the two Borel summations (above and below the positive real axis) have a different imaginary part because they are actually two different analytical continuations of a function with domain in a Riemann surface. The directional Laplace transform in \Eqref{eq_connection_directional_laplace} is also considered in \cite{AnicetoBasarSchiappa} with similar interpretations.

%%%%%%%%%%%%%%%%%%%%%%%%%%%%%%%%%%%%%%%%%%%%%%%%%%%%%%%%%%%%%%%%%%%%%%%%%%%%%%%%%%%%%%%%%%%%%%%%%%%%

\section{Nevanlinna's theorem}
\label{sec_connection_nevanlinna}
It is impossible in general to reconstruct a function solely from the information contained in its power-like asymptotic expansion. For example, two functions that differ by an exponentially suppressed term have the same asymptotic expansion. Nevanlinna's theorem provides sufficient conditions which forbid the presence of exponential corrections hidden beyond the asymptotic expansion and, in those circumstances, the function can be in fact recovered from this expansion by the process of Borel summation. Let us first state the theorem. The proof is sketched in \cite{Sokal}.

We review the theorem and its discussion because it contains relevant observations that will prove useful in later sections.

\begin{theorem}[Nevanlinna's theorem]
\label{theorem_connection_nevanlinna}
\begin{enumerate}[label=(\roman*),wide]
\item \label{it_connection_nevanlinna(a)}
Let $f$ be an analytic function in $\Re(z)>A$ and satisfy there
\begin{equation}
f(z) = \sum_{n=0}^{N-1} \frac{a_n}{z^{n+1}} + R_N(z)
\label{eq_connection_series}
\end{equation}
with
\begin{equation}
\left| R_N(z) \right| \le L(N+1)! \big( \sigma/|z| \big)^{N+1},
\label{eq_connection_remainder}
\end{equation}
where $L > 0$ and $\sigma > 0$. (We say $f$ admits $\sum a_n/z^{n+1}$ as a uniform 1-Gevrey asymptotic expansion in $\Re(z)>A$).

Under the above hypothesis, the series
\begin{equation}
B(\zeta) = \sum_{n\ge 0} \frac{a_n\zeta^n}{n!}
\end{equation}
converges in $|\zeta|<1/\sigma$ and has an analytic continuation to the strip $S_\sigma = \bigcup_{\zeta_0 \in \mathbb{R}^+} \! D(\zeta_0, 1/\sigma)$, where $D(\zeta_0,r)$ is an open disc of centre $\zeta_0$ and radius $r$.\footnote{The bound in \Eqref{eq_connection_remainder} indicates that the coefficients $a_n$ are, at most, factorially divergent, but there is no restriction on their phase. In particular, they could have fixed sign, which would then be incompatible with the fact that $B$ has no singularities in the positive real axis. In \Exref{example_connection_E1}, we will clarify this apparent contradiction and understand that the condition that $f$ is analytic in $\Re(z)>A$ implicitly forces that the $a_n$ cannot be of fixed sign.} Furthermore,
\begin{equation}
| B(\zeta) | \le K e^{A|\zeta|}\,,
\label{eq_connection_BT_bound}
\end{equation}
with $K>0$, in every strip $S_{\sigma'}$ with $\sigma' > \sigma$, and $f$ can be recovered from the Laplace transform
\begin{equation}
f(z) = \int_0^\infty \dd\zeta\, e^{-z\zeta}\, B(\zeta)\,,\quad \Re(z)>A\,.
\label{eq_connection_Laplace}
\end{equation}

\item \label{it_connection_nevanlinna(b)}
If $B(\zeta)$ is analytic in the strip $S_{\sigma'}$ (for all $\sigma' > \sigma$) and there satisfies the bound of \Eqref{eq_connection_BT_bound}, then the function $f$ defined by \Eqref{eq_connection_Laplace} is analytic in $\Re(z) > A$ and admits $\sum B^{(n)}(0)/z^{n+1}$ as a uniform 1-Gevrey asymptotic expansion in $\Re(z)>a$, for any $a>A$.
\end{enumerate}
\end{theorem}

Part \ref{it_connection_nevanlinna(a)} of the theorem gives sufficient conditions under which $f$ is uniquely recovered from the coefficients $a_n$ and further presents a particular way to do so: through the Borel summation of \Eqref{eq_connection_Laplace}.

Part \ref{it_connection_nevanlinna(b)} specifies that the sufficient conditions of part \ref{it_connection_nevanlinna(a)} are also \textit{necessary} in the following sense. If \Eqref{eq_connection_series} and \Eqref{eq_connection_remainder} are not satisfied at least in a region of the type $\Re(z)>a$, then $f$ cannot be recovered from its asymptotic expansion using the Borel summation in \Eqref{eq_connection_Laplace} (although that does not mean $f$ cannot be uniquely recovered through other methods, as we will see in \Exref{example_connection_E1}).

In essence, Nevanlinna's theorem states that, if $f$ satisfies \Eqref{eq_connection_series} and \Eqref{eq_connection_remainder} in $\Re(z)>A$, then the remainder $R_N$ cannot contain any exponential corrections. For example, if $R_N(z)$ contained $e^{-z}$, then we would see this term when taking the limit $|z|\rightarrow \infty$ with $z$ in parallel to the imaginary axis (always keeping $z$ inside $\Re(z)>A$). Indeed, $|e^{-z}| = e^{-\Re(z)}$ would not approach 0 for large $z$, because $\Re(z)$ would go to a constant. Additional discussion regarding this point can be found in \cite{Sokal}. 

\begin{example}
\label{example_connection_E1}
If the hypotheses in part \ref{it_connection_nevanlinna(a)} are satisfied, one of the implications is that the Borel transform of $f$ must be analytic in some strip $S_\sigma$. Let us discuss the following function, defined in terms of the exponential integral $E_1$:
\begin{equation}
f(z) = -e^{-z}E_1(-z) \sim \sum_{n\ge 0} \frac{n!}{z^{n+1}}\,,\quad |z|\rightarrow \infty\,,
\label{eq_connection_f_def}
\end{equation}
whose Borel transform is $B(\zeta)=1/(1-\zeta)$, with an explicit singularity at $1 \in S_\sigma$. This implies either $f$ is not analytic in any of the regions $\Re(z) > A$ or the bound of \Eqref{eq_connection_remainder} fails there.

The exponential integral $E_1$ is a multivalued function that can be written as
\begin{equation}
E_1(z) = -\gamma - \log(z) - \sum_{k\ge 1} \frac{(-z)^k}{k!\,k}\,.
\label{eq_connection_E1}
\end{equation}
Because the series on the right defines an entire function, this expression shows that $f$ has a logarithmic singularity at $z=0$.

The branch cut of the logarithm is conventionally placed along $\mathbb{R}^-$. Thus, $f$ has a branch cut along $\mathbb{R}^+$. It is obvious in this case that $f$ is not analytic in $\Re(z) > A$ for any $A$. However, choosing a different branch for the logarithm, the branch cut may be placed along $\mathbb{R}^-$. Namely, consider the function
\begin{equation}
\begin{cases}
-e^{-z}E_1(-z) & \text{if } \Im(z) \le 0\\
-e^{-z} (E_1(-z) - 2\pi i) & \text{if } \Im(z) > 0\,.
\end{cases}
\label{eq_connection_F_cut}
\end{equation}
This function has the same Borel transform $B$ as $f$ (because the singularities of $f$ did not change), but now its branch cut stretches along $\mathbb{R}^-$. In moving the cut, we have introduced an exponential term that is not suppressed along $i\mathbb{R}^+$. Therefore, by making the function analytic in $\Re(z) > A$, the remainder $R_N$ no longer satisfies the bound in \Eqref{eq_connection_remainder}.

Still, $f$ can be in fact uniquely recovered from its asymptotic expansion, in the sense that $f$ is the only function that has the asymptotic expansion $\sum n!/z^{n+1}$ uniformly valid in $\Re(z)<0$ (compared to $\Re(z)>0$). We need a slight modification of \Theoref{theorem_connection_nevanlinna}. Instead of the region $\Re(z)>A$ in part \ref{it_connection_nevanlinna(a)} of the theorem, we consider the generalised region $\Re\big( ze^{i\theta} \big)>A$, which is the half-plane bisected by the half-line $e^{-i\theta}\mathbb{R}^+$ and whose boundary is at a distance $A$ from 0 (see \Figref{fig_connection_nevanlinna}). In addition, the strip where $B$ is analytic and satisfies the bound of \Eqref{eq_connection_BT_bound} is replaced by $S_\sigma(\theta) = \bigcup_{\zeta_0 \in e^{i\theta}\mathbb{R}^+} \! D(\zeta_0,1/\sigma)$. Then, $f$ can be recovered from the directional Laplace transform
\begin{equation}
f(z) = \int_0^{\infty e^{i\theta}}\! \dd\zeta \, e^{-z\zeta}B(\zeta)\,, \quad \Re\big( ze^{i\theta} \big)>A\,.
\end{equation}
Part \ref{it_connection_nevanlinna(b)} of the theorem may be modified in the same way.
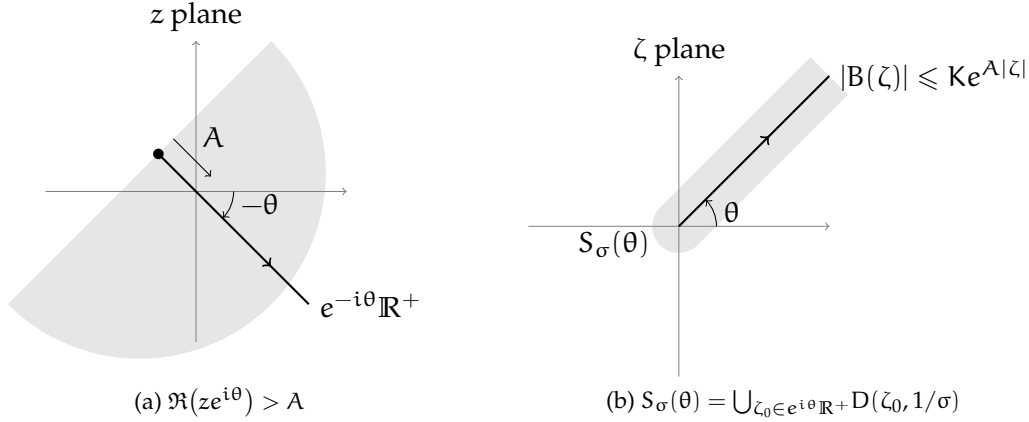
\begin{figure}
\centering
\subfloat[$\Re\big( ze^{i\theta} \big)>A$\label{fig_connection_nevanlinna_(a)}]{
\begin{tikzpicture}[scale=1]
\fill[fill=gray!20!white] (1,2) -- (-2.5,-1.5) arc (-135:45:2.475);

\draw[help lines,->] (-2,0) -- (2,0) coordinate (xaxis);
\draw[help lines,->] (0,-2) -- (0,2) coordinate (yaxis);

%\draw (1,2) -- (-2.5,-1.5);

\path[draw,line width=0.8pt,decoration={markings, mark=at position 0.75 with {\arrow{>}}},
postaction=decorate] (-0.5,0.5) -- (1.5,-1.5) node[right]{$e^{-i\theta}\mathbb{R}^+$};

\draw[->] (0.5,0) arc (0:-44:0.5) node[midway,right]{$-\theta$};

\fill (-0.5,0.5) circle[radius=2pt];

\draw[->] (-0.3,0.7) -- node[midway,above right]{$A$} (0.2,0.2); 

\node[above] at (yaxis) {$z$ plane};
\end{tikzpicture}
}
\hspace{1cm}
\subfloat[$S_\sigma(\theta) = \bigcup_{\zeta_0 \in e^{i\theta}\mathbb{R}^+} \! D(\zeta_0,1/\sigma)$\label{fig_connection_nevanlinna_(b)}]{
\begin{tikzpicture}[scale=1]
\fill[fill=gray!20!white] (1.75,2.25) -- (-0.25,0.25) arc(135:315:0.353553) node[midway,left]{$S_\sigma(\theta)$} -- (2.25,1.75);

\draw[help lines,->] (-2,0) -- (2,0) coordinate (xaxis);
\draw[help lines,->] (0,-2) -- (0,2) coordinate (yaxis);

\path[draw,line width=0.8pt,decoration={markings, mark=at position 0.6 with {\arrow{>}}},
postaction=decorate] (0,0) -- (2,2) node[right]{$|B(\zeta)|\le K e^{A|\zeta|}$};

\draw[->] (0.5,0) arc (0:44:0.5) node[midway,right]{$\theta$};

\node[above] at (yaxis) {$\zeta$ plane};
\end{tikzpicture}
}
\caption{Regions where the generalisation of \Theoref{theorem_connection_nevanlinna} applies.}
\label{fig_connection_nevanlinna}
\end{figure}

For the example at hand, we might consider $\theta = \pi$. The function $f$ is analytic in $\Re(z)<0$ and admits $\sum n!/z^{n+1}$ as a uniform 1-Gevrey asymptotic expansion in that region. Thus $f$ can be uniquely recovered from the coefficients $a_n = n!$ in $\Re(z)<0$ through Borel summation along the direction $\theta = \pi$.
\end{example}

This example shows that it is not necessary that a function satisfies the hypothesis of part \ref{it_connection_nevanlinna(a)} of Nevanlinna's theorem in $\Re(z)>A$, but it is enough if they are satisfied in some half-plane $\Re\big( ze^{i\theta} \big)>A$.

The example also shows that the situations between $f(z)$ and $f(-z)$ are symmetric. For $f(-z)$, the Borel transform is $B(-\zeta)$, which has a pole at $\zeta=-1$, so it does not interfere with the standard Borel summation. For $f(z)$, even if the pole at $\zeta=1$ interferes with the summation, we may just change the direction of summation. In both cases, the function may be uniquely recovered from its asymptotic expansions, as we would naturally expect.

As a final remark, we notice that $f(z)$ in \Eqref{eq_connection_f_def} has an imaginary exponentially small part for $z>0$, coming from the logarithm in \Eqref{eq_connection_E1}. This imaginary part is ambiguous and can also be traced to the presence of the pole at $\zeta=1$. The asymptotic expansions of path integrals in powers of the coupling $g=1/z$ sometimes are non-alternating, as in \Eqref{eq_connection_f_def}. Thus the Borel sums of these expansions also have imaginary exponentially small parts for $z>0$. Nevertheless, we expect that path integrals are real for positive coupling ($z>0$). Therefore we will always need exponential corrections to cancel those imaginary parts. In particular, this means that these path integrals will never satisfy the conditions of Nevanlinna's theorem or its generalisation.

%%%%%%%%%%%%%%%%%%%%%%%%%%%%%%%%%%%%%%%%%%%%%%%%%%%%%%%%%%%%%%%%%%%%%%%%%%%%%%%%%%%%%%%%%%%%%%%%%%%%

\section{The resurgent connection, a first approach by dispersion relations}
\label{sec_connection_heuristic}
In the present section, we will review the derivation of the resurgent connection developed by \cite{BenderWu}, which is based on the existence of a dispersion relation. This part will serve as an introduction for \Secref{sec_connection_formal}, where we will give precise conditions that guarantee the resurgent connection.

Given a function $f$ analytic in $\mathbb{C}\setminus\mathbb{R}^-$, we assume that
\begin{equation}
f(z) \sim \sum_{n\ge 0}\frac{a_n}{z^{n+1}}\,, \quad z\rightarrow +\infty\,,
\label{eq_connection_f_beh}
\end{equation}
and
\begin{equation}
\disc{f}(z) \sim 2i\, b_0\, e^{-Sz}(-z)^{\alpha-1}\,, \quad z\rightarrow -\infty\,,
\label{eq_connection_disc_beh}
\end{equation}
where $\disc{f}(z) = f(z+i0) - f(z-i0)$ with $z<0$. The resurgent connection is the relation between the large $-z$ behaviour of $\disc{f}$ and the large order behaviour of the coefficients $a_n$.

If $f$ is a path integral, this is in fact a connection between large order perturbative physics and low order non-perturbative physics. The coefficients $a_n$ in \Eqref{eq_connection_f_beh} are computed from quantum fluctuations around the trivial saddle point (saddle points with zero action), while the coefficient $b_0$ in \Eqref{eq_connection_disc_beh} is computed from fluctuations around non-trivial saddle points (non-zero action).\footnote{For example, in \cite{CollinsSoper}, where the ground energy of the anharmonic oscillator is discussed, $b_0$ is computed in this way. In \cite[Sec.~2.3]{Marino2015}, this computation is carried out for a 0-dimensional quartic interaction.} It is in this sense that the connection is \guillemotleft resurgent\guillemotright{}.

We note that Lipatov's method \cite{Lipatov} also describes a connection between the large order behaviour of the $a_n$ and the non-trivial saddle points of the action (although in this case, the connection exists with no mention to the discontinuity of $f$ at all). See \cite{Suslov} for an illustration of this method on different field models.

The power of the resurgent connection is that a single diagram, encoding $b_0$, is enough to determine the values of the $a_n$ for large $n$, an information that would require the computation of an infinite number of diagrams otherwise.

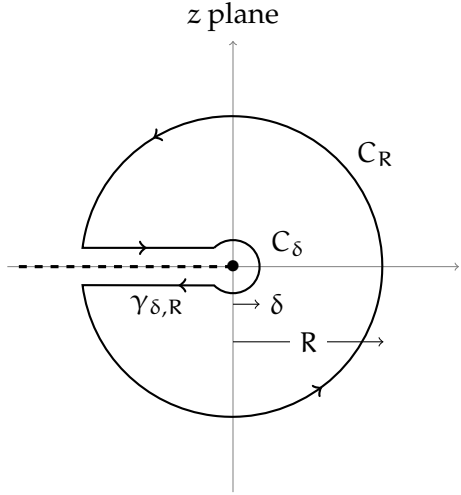
\begin{figure}
\centering
\begin{tikzpicture}[decoration={markings,
mark=at position 1cm with {\arrow[line width=1pt]{<}},
mark=at position 3.65cm with {\arrow[line width=1pt]{<}},
mark=at position 9.75cm with {\arrow[line width=1pt]{<}},
mark=at position 15.2cm with {\arrow[line width=1pt]{<}}
}
]

% The axes
\draw[help lines,->] (-3,0) -- (3,0) coordinate (xaxis);
\draw[help lines,->] (0,-3) -- (0,3) coordinate (yaxis);

% The path
\path[draw,line width=0.8pt,postaction=decorate] (-0.25,0.25) -- (-2,0.25) arc (172.875:-172.875:2) -- (-0.25,-0.25) arc (-135:135:0.353553) node[midway,above right]{$C_\delta$};

\draw[->] (0,-1) -- (2,-1) node[midway,fill=white] {$R$};
\draw[->] (0,-0.5) -- (0.353553,-0.5) node[right] {$\delta$};

\draw[dashed,very thick] (0,0) node{$\bullet$} -- (-2.95,0);

% The labels
\node[above] at (yaxis) {$z$ plane};
\node[below] at (-1,-0.25) {$\gamma_{\delta,R}$};
\node[above,right] at (1.5,1.5) {$C_R$};
\end{tikzpicture}
\caption{Closed path for a dispersion relation. The width of the path $\gamma_{\delta,R}$ around $\mathbb{R}^-$ is already taken to 0.}
\label{fig_connection_dispersion}
\end{figure}

To determine the exact resurgent connection, we consider the closed path in \Figref{fig_connection_dispersion}. Using the residue theorem, we have
\begin{equation}
f(z) = -\frac{1}{2\pi i} \int_\delta^R \dd w\, \frac{\disc{f}(-w)}{w+z} + I_\delta(z) + \mathcal{I}_R(z)\,,
\label{eq_connection_dispersion_I+I}
\end{equation}
where $I_\delta(z)$ and $\mathcal{I}_R(z)$ are the integrals of $f(w)/(w-z)/(2\pi i)$ around $C_\delta$ and $C_R$, respectively. The width of $\gamma_{\delta,R}$ is already taken to 0 and so the integral along this path can be written as an integral in $[\delta,R]$ of the discontinuity of $f$.

First, let us assume that $I_\delta(z)$ does not contribute to the large order behaviour of the $a_n$. In essence, what happens at small $z$ (around the circle $C_\delta$) should be independent of the behaviour at large $z$ (encoded in the coefficients $a_n$). This argument is heuristic at this point, but we will formalise it in \Secref{sec_connection_formal}. We note that the hypothesis $I_\delta(z) \rightarrow 0$ when $\delta \rightarrow 0$ was needed in \cite{BenderWu} in order to derive the resurgent connection. We are improving on the original discussion by not demanding any condition on $I_\delta(z)$.

Second, we assume that $|\mathcal{I}_R(z)| \rightarrow 0$ for $R\rightarrow\infty$.\footnote{Under this assumption, we say $f$ satisfies a dispersion relation (regardless of the contribution from $I_\delta$). Also note that \guillemotleft dispersion relation\guillemotright{} is commonly associated with the $q^2$ plane, where $q$ is some relevant momentum. Here we use this term with no regards to the physical interpretation of $z$.} This assumption is essential in the discussion of \cite{BenderWu}. We will show in \Exref{example_connection_exp_correction} below that if this hypothesis fails, then the resurgent connection may not take place in general. Still, in \Secref{sec_connection_median}, we will be able to identify a less restrictive set of functions for which the connection holds even if $\mathcal{I}_R(z)$ does not vanish for large $R$ (these will be the functions obtained from median resummation).

Finally, we invoke the Taylor expansion of $1/(w+z)$ around $w=0$ and integrate term by term in \Eqref{eq_connection_dispersion_I+I}:
\begin{equation}
\begin{aligned}
f(z) - I_\delta(z) &= -\frac{1}{2\pi i} \int_\delta^\infty \dd w\, \frac{\disc{f}(-w)}{w+z}\\
&= -\frac{1}{2\pi i} \int_\delta^\infty \dd w\, \disc{f}(-w)\, \frac{1}{z} \sum_{n\ge 0} \left( - \frac{w}{z} \right)^n \\
&\sim \sum_{n\ge 0} \frac{1}{z^{n+1}}\left[ \frac{(-1)^{n+1}}{2\pi i} \int_\delta^\infty \dd w\, w^n\, \disc{f}(-w) \right].
\end{aligned}
\label{eq_connection_integrate_disc}
\end{equation}
Notice that the Taylor expansion of $1/(w+z)$ should only be valid inside the disc of convergence $|w|<|z|$, but the line of integration stretches much beyond this region for any finite $z$. This is the typical situation where integrating term by term yields a divergent asymptotic expansion, rather than a convergent series.

Comparing the last line in \Eqref{eq_connection_integrate_disc}, with the asymptotic expansion $f(z) \sim \sum a_n/z^{n+1}$, we already conclude that
\begin{equation}
a_n \simeq \frac{(-1)^{n+1}}{2\pi i} \int_\delta^\infty \dd w\, w^n \disc{f}(-w)\,,
\label{eq_connection_dispersion_an}
\end{equation}
and the equality is exact up to corrections coming from $I_\delta(z)$.

As we are only interested in the high order behaviour of the $a_n$, we ignore the contribution form $I_\delta(z)$ and choose $\delta$ large enough so that we may replace $\disc{f}(-z)$ by its behaviour at large $-z$, quoted in \Eqref{eq_connection_disc_beh}. After integration we obtain
\begin{equation}
a_n \sim \frac{(-1)^{n+1}}{\pi}\frac{\Gamma(n+\alpha)}{(-S)^{n+\alpha}}\, b_0\,, \quad n\rightarrow \infty\,.
\label{eq_connection_an_beh}
\end{equation}
This concludes the derivation of the resurgent connection.

We would like to warn the reader that there is a caveat with this derivation of \Eqref{eq_connection_an_beh}. In \Appref{sec_connection_ordering}, we discuss an example where all the assumptions in the present section are satisfied, but even then there is no correspondence between \Eqref{eq_connection_disc_beh} and \Eqref{eq_connection_an_beh}. In fact, the reason this derivations is not complete is because we did not keep track of the error between the true $a_n$ and the approximation in \Eqref{eq_connection_an_beh}. We will take care of this issue in \Secref{sec_connection_formal}.

In the following, we discuss a simple example to understand the importance of the assumption that $\mathcal{I}_R(z)$ vanishes at large $R$.
\begin{example}
\label{example_connection_exp_correction}
Consider the function
\begin{equation}
g(z) = -e^{-z}\big[ E_1(-z) + \log(-z) \big]\,,
\end{equation}
which is the same function as in \Exref{example_connection_E1}, but with an additional exponential term. From \Eqref{eq_connection_E1}, it is easy to check that the function of this example is entire. In particular, this means its discontinuity is 0.

An asymptotic expansion for $g$ is given by
\begin{equation}
g(z) \sim \sum_{n\ge 0} \frac{n!}{z^{n+1}}\,, \quad \Re(z)>0\,,
\label{eq_connection_expansion_F}
\end{equation}
which is the same expansion as in \Eqref{eq_connection_f_def}, but the region of validity cannot be extended past the imaginary axis. This is because the exponential term $-e^{-z}\log(-z)$ becomes enhanced in $\Re(z)<0$. In fact, the expansion is not uniformly valid in $\Re(z)>0$, because the modulus of the exponential term goes like $\log|z|$ along lines parallel to the imaginary axis.

The function $g$ does not feature a resurgent connection. If it did, given the 0 discontinuity (smaller than $e^{-Sz}$ at large $-z$ for all $S<0$), the coefficients $a_n$ should be smaller than $n!/(-S)^n$ for all $S<0$, which clearly is not the case.

Of course, this function fails to realise the assumptions that we demanded in the previous derivation. In particular, it fails the hypothesis that $\mathcal{I}_R(z)$ vanishes at large $R$, due to the presence of the exponential term $-e^{-z}\log(-z)$.
\end{example}

One might think that $f$ will always satisfy a dispersion relation provided we perform enough subtractions. This is correct up to some point. Leading terms in the asymptotic expansion like $z^n$ or $\log^n(z)$ (with $n\ge 0$) can be eliminated until the subtracted function vanishes for $|z| \rightarrow \infty$, so it satisfies a dispersion relation. But exponential corrections hidden beyond the asymptotic expansion cannot be dealt in the same way, so they will always spoil the dispersion relation.

%A lesson emerges from the above example. If $f$ does not satisfy a dispersion relation because its asymptotic expansion contains leading terms (like $z^n$ or $\log^n(z)$ with $n\ge 0$), the resurgent connection will be satisfied after the subtraction of those leading terms. In particular, this is important when $f$ is a correlator ($z=q^2$), which satisfy dispersion relations in $q^2$ after subtractions.

%Instead, if the dispersion relation is not satisfied because of exponential corrections hidden beyond all terms of the asymptotic expansion, then we cannot bypass this problem as in the previous paragraph.

In spite of this, we notice that the resurgent connection can still take place if the exponential corrections make no contribution to the discontinuity of $f$. As we will see in \Secref{sec_connection_median}, this last observation will be central to the generalisation of the resurgent connection beyond functions that satisfy a dispersion relation. In this sense, our discussion will generalise that of \cite{BenderWu}.

%%%%%%%%%%%%%%%%%%%%%%%%%%%%%%%%%%%%%%%%%%%%%%%%%%%%%%%%%%%%%%%%%%%%%%%%%%%%%%%%%%%%%%%%%%%%%%%%%%%%

\section{The resurgent connection, formal statements}
\label{sec_connection_formal}
Before presenting the formal statements of the resurgent connection, we will develop some intuition by discussing the particular example below.

\begin{example}
\label{example_connection_resurgence}
The function $f(z) = e^z E_1(z)$ has the discontinuity
\begin{equation}
\disc{f}(z) = -2\pi i e^{z}\,, \quad z<0\,,
\end{equation}
which can be computed from the logarithm in \Eqref{eq_connection_E1}.

The inverse Laplace transform of $f$ is
\begin{equation}
B(\zeta) = -\frac{1}{2\pi i} \int_{\gamma_{0,\infty}} \! \dd z\, e^{z\zeta}\, f(z) = \frac{1}{2\pi i} \int_0^{-\infty} \! \dd z\, e^{z\zeta}\, \disc{f}(z) = \frac{1}{\zeta+1}\,.
\end{equation}
Here we have started from the definition in \Eqref{eq_connection_borel} and deformed the path $\mathcal{C}_a$ into $\gamma_{0,\infty}$. Note that this deformation is possible because $f$ goes to 0 for large $|z|$ in the region $\Re(z)\le a$.

\begin{sloppypar}
Using the integral representation of $E_1$, one can analytically check that the Laplace transform of $B$ recovers $f$. Thus, in this case,
\begin{equation}
f(z) = \int_0^\infty \dd\zeta\, e^{-z\zeta}\, B(\zeta) \sim \sum_{n\ge 0}\frac{(-1)^n n!}{z^{n+1}}\,, \quad \Re(z)>0\,,
\end{equation}
where the asymptotic expansion is obtained from part \ref{it_connection_nevanlinna(b)} of Nevanlinna's theorem, with $B^{(n)}(0)=(-1)^n n!$.
\end{sloppypar}

This is an explicit verification of the resurgent connection, where the discontinuity in \Eqref{eq_connection_disc_beh} fixes the singularities of $B$ and, in turn, the singularities determine the large order behaviour in \Eqref{eq_connection_an_beh} of the coefficients $a_n$ (in this case, the result is exact). This time we have used the Borel framework as the main tool of the derivation, rather than a dispersion relation.

In addition, we now check that the converse resurgent connection also holds. That is, the large order behaviour of the coefficients $a_n$ fixes the singularities in $B$ and, in turn, the singularities determine the discontinuity of $f$.

We consider a function $f$ whose asymptotic expansion is $f(z) \sim \sum (-1)^n n!/z^{n+1}$, uniformly valid in $\Re(z)>0$. In this case, from part \ref{it_connection_nevanlinna(a)} of Nevanlinna's theorem,
\begin{equation}
\begin{array}{ll}
\displaystyle B(\zeta) = \sum_{n\ge 0} \frac{a_n \zeta^n}{n!} = \frac{1}{\zeta + 1}\,; & \displaystyle f(z) = \int_0^\infty \dd\zeta\, e^{-z\zeta}\, B(\zeta)\,, \quad \Re(z)>0\,.
\label{eq_connection_ftheta0}
\end{array}
\end{equation}
It is easy to check that $|B(\zeta)| \le K$ in $|\zeta|>R$, for some $K$, $R>0$.

We consider the family of functions
\begin{equation}
f_\theta(z) = \int_0^{\infty e^{i\theta}} \! \dd\zeta\, e^{-z\zeta}\, B(\zeta)\,, \quad \Re\big( ze^{i\theta}\big) > 0\,,
\label{eq_connection_ftheta}
\end{equation}
and verify that, for $\theta_+$, $\theta_-$ such that $0\le |\theta_+ - \theta_-| < \pi$, the closed path in \Figref{fig_connection_ftheta(a)} yields
\begin{equation}
f_{\theta_\plus}(z) - f_{\theta_\minus}(z) = \lim_{R\rightarrow \infty}\int_{C_R} \! \dd\zeta\, e^{-z\zeta}\, B(\zeta) = 0\,, \quad \Re\big( ze^{i\theta_\plus} \big)>0 \cap \Re\big( ze^{i\theta_\minus} \big)>0\,.
\label{eq_connection_ftheta=}
\end{equation}
For any $z$ in the intersection of the two half-planes, the integrand is exponentially suppressed all along $C_R$, thus the integral vanishes. This result can be formalized using the bound on $|B|$.

\begin{figure}
\centering
\subfloat[\label{fig_connection_ftheta(a)}]{
\begin{tikzpicture}[decoration={markings,
mark=at position 0.15 with {\arrow[line width=1pt]{<}},
mark=at position 0.48 with {\arrow[line width=1pt]{<}},
mark=at position 0.85 with {\arrow[line width=1pt]{<}}
}
]

% The axes
\draw[help lines,->] (-3,0) -- (3,0) coordinate (xaxis);
\draw[help lines,->] (0,-3) -- (0,3) coordinate (yaxis);

% The path
\path[draw,line width=0.8pt,postaction={decorate}] (0,0) -- (2,2) node[above] {$-f_{\theta_\plus}$} arc (45:-45:2.82843) node[midway, above left] {$C_R$} node[below] {$f_{\theta_\minus}$} -- (0,0);
\draw[|->] (0.5,0) arc (0:44:0.5) node[below=1mm, right=1mm]{$\theta_+$};
\draw[|->] (0.5,0) arc (0:-44:0.5) node[above=1mm, right=1mm]{$\theta_-$};

% The labels
\node[above] at (yaxis) {$\zeta$ plane};
\node at (-1,0) {$\bullet$};
\node[below] at (-1,0) {$-1$};
\end{tikzpicture}
}
\hspace{1cm}
\subfloat[\label{fig_connection_ftheta(b)}]{
\begin{tikzpicture}[decoration={markings,
mark=at position 0.15 with {\arrow[line width=1pt]{>}},
mark=at position 0.48 with {\arrow[line width=1pt]{>}},
mark=at position 0.85 with {\arrow[line width=1pt]{>}}
}
]

% The axes
\draw[help lines,->] (-3,0) -- (3,0) coordinate (xaxis);
\draw[help lines,->] (0,-3) -- (0,3) coordinate (yaxis);

% The path
\path[draw,line width=0.8pt,postaction={decorate}] (0,0) -- (-2,2) node[above] {$f_{\theta_\plus}$} arc (135:225:2.82843) node[midway, above right] {$C_R$} node[below] {$-f_{\theta_\minus}$} -- (0,0);
\draw[|->] (0.5,0) arc (0:134:0.5) node[midway, right=1mm]{$\theta_+$};
\draw[|->] (0.5,0) arc (0:-134:0.5) node[midway, right=1mm]{$\theta_-$};

% The labels
\node[above] at (yaxis) {$\zeta$ plane};
\node at (-1,0) {$\bullet$};
\node[below] at (-1,0) {$-1$};
\end{tikzpicture}
}
\caption{Relationships between the family of functions $f_{\theta}$ defined in \Eqref{eq_connection_ftheta}.}
\label{fig_connection_ftheta}
\end{figure}
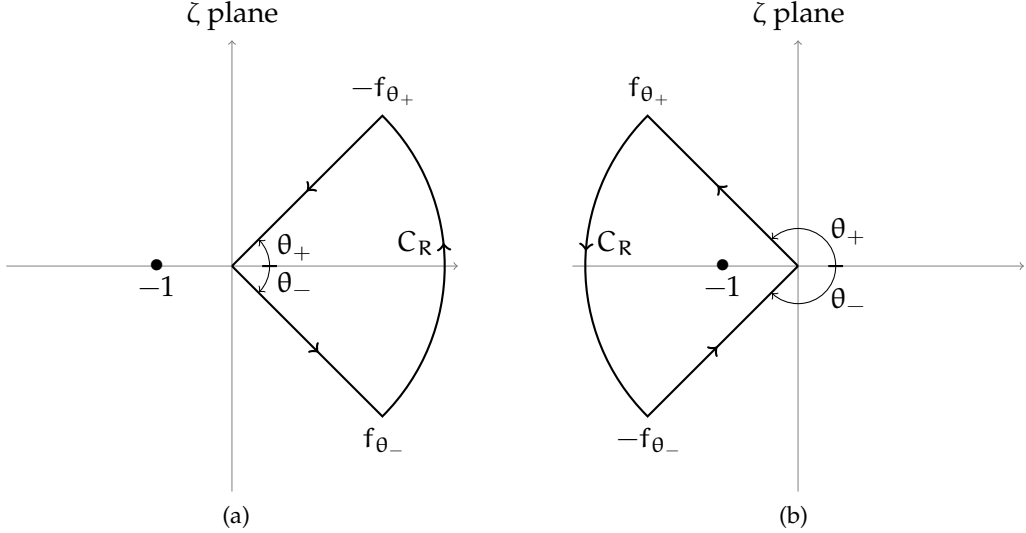

From \Eqref{eq_connection_ftheta=}, one understands that the family of functions $f_\theta$ are just analytical continuations of the same function $f=f_0$ to different regions of the complex plane.

On the other hand, when we consider the two directions in \Figref{fig_connection_ftheta(b)}, the integral along $C_R$ again vanishes, but we now obtain a contribution from the residue of the integrand at $\zeta=-1$:
\begin{equation}
f_{\theta_\minus}(z) - f_{\theta_\plus}(z) = -2\pi i e^z\,, \quad \Re\big( ze^{i\theta_\plus} \big)>0 \cap \Re\big( ze^{i\theta_\minus} \big)>0\,.
\label{eq_connection_jump}
\end{equation}
This result is easily understood in the Riemann surface of $f$, plotted in \Figref{fig_connection_riemann1}. The two directions $\theta_+$, $\theta_-$ probe different regions of the Riemann surface (depicted in gray in the figure) with the same projection in the complex plane. The discontinuity of $f$ is the jump between the two regions,\footnote{From now on, we define $\disc{f}$ in this way, rather than as the limit $\disc{f}(z) = f(z+i0) - f(z-i0)$. Note that, using this redefinition, $\disc{f}$ is itself an analytic function and, as such, it admits analytical continuations in its own Riemann surface. This will be relevant in \Secref{sec_connection_median}.} which exactly corresponds to the difference in \Eqref{eq_connection_jump}. That is, $\disc{f}(z) = -2\pi i e^z$, as we expected.

\begin{figure}
\centering
\begin{overpic}[width=0.5\textwidth]{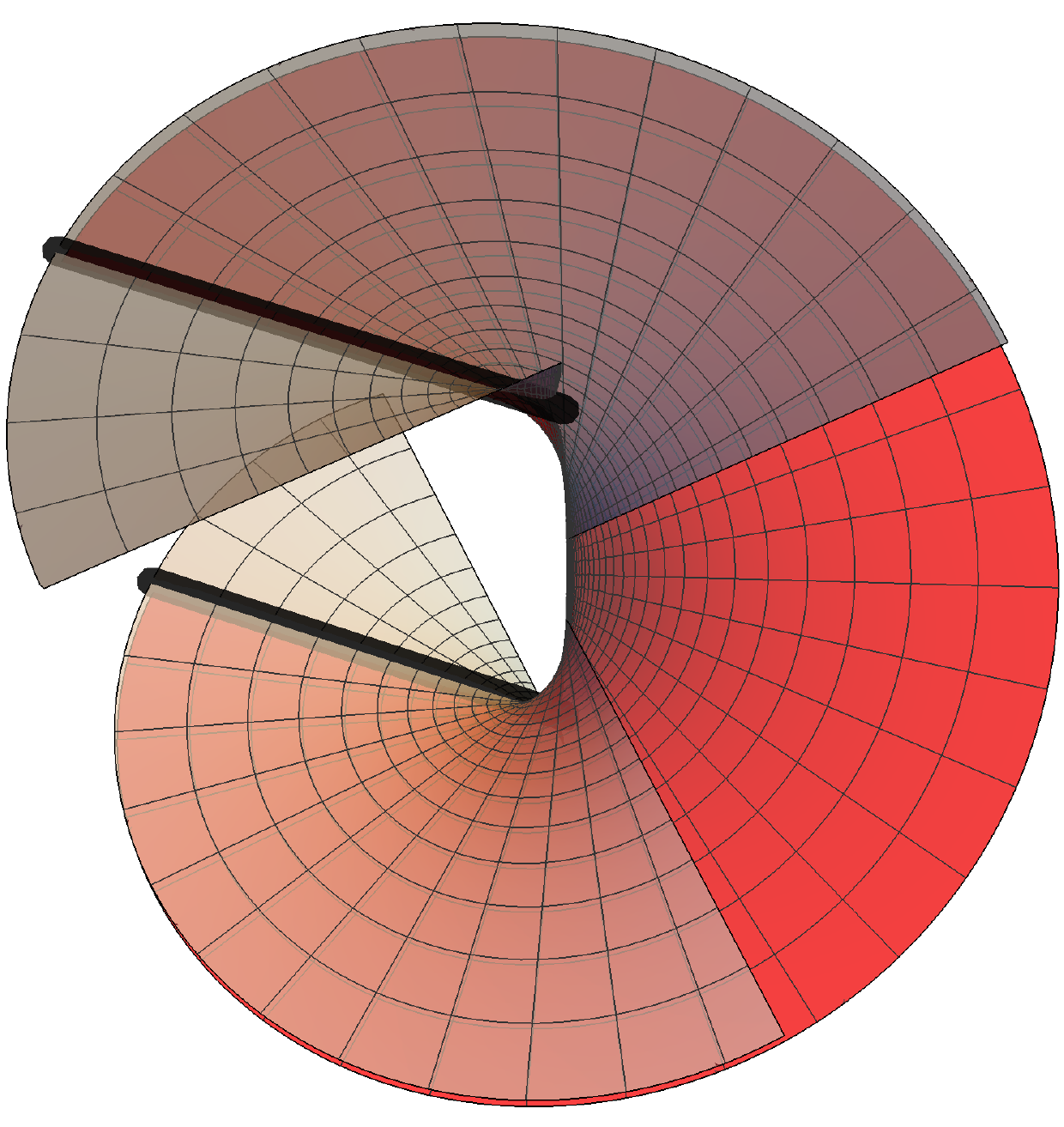}
\put (98,35) {$\Re(z)>0$}
\put (40,81) {$f_{\theta_\minus}$}
\put (30,20) {$f_{\theta_\plus}$}
\put (-20,80) {$\Re(z)<0$}
\linethickness{0.75pt}
\put (25,45) {\vector(0,1){26}}
\put (27,57) {$\disc{f}(z)$}
\end{overpic}
\caption{Riemann surface of $f$, defined in \Eqref{eq_connection_ftheta0}. The regions of analytical continuation given by $f_{\theta_\plus}$ and $f_{\theta_\minus}$ are depicted in grey.}
\label{fig_connection_riemann1}
\end{figure}
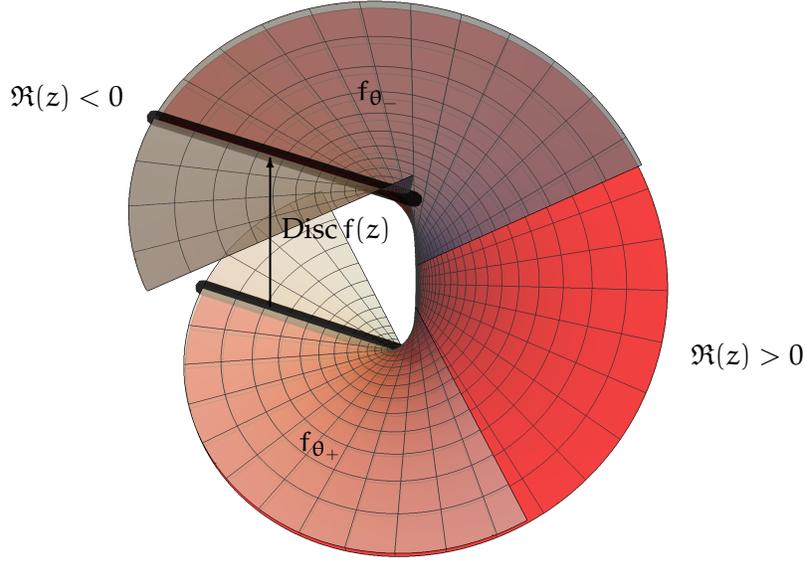
\end{example}

Let us argue why it was somehow expected that the discontinuity of a function and the coefficients $a_n$ are related through the singularities in the Borel plane. An heuristic argument by 't Hooft shows that instanton singularities in the Borel plane are determined by the value of the action that each non-trivial saddle point attains \cite{tHooft} (also see \cite[Sec.~4.6]{Marino2015} for a review).

In quantum mechanics, we know that the discontinuity of the path integral in \Eqref{eq_connection_intro_E_path_2} is computed from the non-trivial saddle points in the action $S$. Simultaneously, these saddle points also determine the position of the singularities in the Borel plane by the 't Hooft argument. Thus, we conclude that the discontinuity is related to the position of the singularities.

%Actually, we can now give a better ground to the 't Hooft argument by reversing the reasoning. If the discontinuity in the path integral has the exponential behaviour $e^{Sz}$ for large $-z$, then its Borel transform has a singularity at $\zeta=S$, as we have seen in \Exref{example_connection_resurgence}. Simultaneously, the discontinuity can be computed from the non-trivial saddle points $\Phi$ with action $S[\Phi,1/z] = Sz$. Thus, it is clear that these saddle points yield a singularity at $\zeta=S$.

After discussing \Exref{example_connection_resurgence}, we are now in a good position to formally state the resurgent connection.
%Proposition 1
\begin{prop}[Resurgent connection]
\label{prop_resurgence_1}
Let $f$ be an analytic function in $\mathbb{C}\setminus(\mathbb{R}^\minus +A)$ and satisfy $|f(z)| \le |a_0|/|z|$ in $\mathbb{C}$ minus a neighbourhood of $A$. Further assume that
\begin{equation}
\disc{f}(z) = 2i\, b_0\, e^{-Sz}(-z)^{\alpha-1}\left[ 1 + \mathcal{O}\left( \frac{1}{z} \right)\right], \quad z\rightarrow -\infty\,,
\label{eq_connection_disc_prop}
\end{equation}
with $S<0$.\footnote{See below for a generalisation to $S \in \mathbb{C}\setminus\mathbb{R}^+$.} Then the Borel transform 
\begin{equation}
B(\zeta) = \frac{1}{2\pi i} \int_{\mathcal{C}_a}\! \dd z\, e^{z\zeta}\, f(z)\,,
\label{eq_connection_borel_prop}
\end{equation}
with $a>A$, is analytic in $\Re(\zeta)>S$ and is exponentially bounded there by $|B(\zeta)| \le K e^{A\Re(\zeta)}$. Furthermore,
\begin{equation}
f(z) = \int_0^{\infty e^{i\theta}} \! \dd\zeta \, e^{-z\zeta}\, B(\zeta) \sim \sum_{n\ge 0} \frac{a_n}{z^{n+1}}\,, \quad \Re\big( ze^{i\theta} \big) > a\cos(\theta)\,,
\label{eq_connection_laplace_prop}
\end{equation}
(the asymptotic expansion being uniformly valid), with $|\theta|<\pi/2$ and
\begin{equation}
a_n = B^{(n)}(0) = \frac{(-1)^{n+1}}{\pi}\frac{\Gamma(n+\alpha)}{(-S)^{n+\alpha}}\, b_0\left[ 1 + \mathcal{O}\left( \frac{1}{n} \right)\right], \quad n\rightarrow \infty\,.
\label{eq_connection_an_prop}
\end{equation}
\end{prop}

The proof can be found in \Appref{sec_connection_proof_1}. There we repeat the steps in the first part of \Exref{example_connection_resurgence}, but for a general function. First we define the Borel transform in \Eqref{eq_connection_borel_prop} and check, using \Eqref{eq_connection_disc_prop}, that its Taylor coefficients $B^{(n)}(0)$ have the large order behaviour in \Eqref{eq_connection_an_prop}. The tricky part is to prove the equality in \Eqref{eq_connection_laplace_prop}, but once this is done, the validity of the asymptotic expansion $f(z) \sim \sum a_n/z^{n+1}$ is just a consequence of part \ref{it_connection_nevanlinna(b)} of Nevanlinna's theorem.

A lesson that may be learned from \Exref{example_connection_resurgence} and \Propref{prop_resurgence_1} is that the resurgent connection is always satisfied by functions that can be written as a Laplace transform. We will transfer this result to the discussion of \Secref{sec_connection_median}.

%There we repeat the computation in \Eqref{eq_connection_integrate_disc}, but with an important modification. The Borel transform of $f$ might be obtained by replacing inside the integral the factor $1/(w+z)$ by the Borel transform of this factor, given by $e^{w\zeta}$:
%\begin{equation}
%\begin{array}{ccc}
%\displaystyle f(z) = -\frac{1}{2\pi i} \int_\delta^\infty\! \dd w\, \frac{\disc{f}(-w)}{w+z} &  \longmapsto & \displaystyle B(\zeta) = -\frac{1}{2\pi i} \int_\delta^\infty\! \dd w\, e^{w\zeta}\, \disc{f}(-w)\,.
%\end{array}
%\end{equation}
%Then we can Taylor expand $e^{w\zeta}$ around $\zeta=0$. Because this later expansion has infinite radius of convergence, now it is legitimate to commute the Taylor series with the integral. This lets us compute the Taylor coefficients $B^{(n)}(0)$, giving the asymptotic behaviour in \Eqref{eq_connection_an_prop}.

\begin{obs}
\label{obs_connection}
As we already pointed out after \Eqref{eq_connection_an_beh} (also in \Appref{sec_connection_ordering}), the resurgent connection cannot take place if the error $\mathcal{O}(1/z)$ in \Eqref{eq_connection_disc_prop} contains further exponential corrections $e^{-S_1 z}$ with $\Re(S_1)<\Re(S)$, but with $|S_1|<|S|$.

We have two different orderings in the Borel plane. The closest singularity to 0 determines the leading behaviour of the $a_n$. This is an ordering in $|\zeta|$. The singularity with the highest real part determines the leading behaviour of $\disc{f}$. This is an ordering in $\Re(\zeta)$. In fact, this was already observed by \cite{Peris} in the context of the operator product expansion and quark-hadron duality.
\end{obs}

In the assumptions of the proposition, we impose $S<0$ in order to deal with Observation \ref{obs_connection}. The Borel transform $B$ is analytic in $\Re(\zeta)>S$, so the singularities in the Borel plane can only be in the region $\Re(\zeta)\le S$. But, because $S<0$, any of these singularities will always be farther away from the origin than $S$. We could admit $S \in \mathbb{C}\setminus\mathbb{R}^+$ in the proposition as long as we have additional assumptions on the Borel plane that forbid the situation of Observation \ref{obs_connection}. A sketch of the proof for this generalised result can be found in \Appref{sec_connection_proof_1'}.

%It is straightforward to include additional corrections into \Eqref{eq_connection_disc_prop}. These would yield sub-leading corrections to \Eqref{eq_connection_an_prop}. To prove this generalisation, we would include the additional corrections to \Eqref{eq_connection_B(n)} and integrate them explicitly as we did for the leading correction.

%The hypothesis $f(z) = a_0/z + \mathcal{O}(1/z^2)$ can be improved. We see in \Eqref{eq_connection_B_bound} that the term $1/z$ is not suppressed enough, so the integral $\int_{\mathcal{C}_a} \! \dd w\, e^{w\zeta}\!/w$ does not converge absolutely. This in principle would be needed to check that the the integral in \Eqref{eq_connection_B=f} converges absolutely and, thus, the order of integration can be interchanged.

%However, while having absolute convergence of the double integral is sufficient, it is not necessary. In the proof we have explicitly verified that the integrals commute for the term $1/z$, even with no absolute convergence.

%One can check that the same property is also true for any term $1/z^\lambda$, with $\lambda>0$. Thus, we might admit $f(z) = a_{\lambda-1}/z^{\lambda} + \mathcal{O}\big(1/z^{\lambda+1}\big)$ instead of the original assumption. This is precisely the behaviour required for a dispersion relation on $f$.

%Proposition 2
In the following, we also present a kind of \guillemotleft converse\guillemotright{}  to \Propref{prop_resurgence_1}. In this case we make no mention to the asymptotic expansion of $f$, but rather we make assumptions directly over the Borel transform. Of course, if $f$ has the asymptotic expansion $\sum a_n/{z^{n+1}}$, then the large order behaviour of the $a_n$ determine the singularities of $B$. It is in this sense that the proposition below is the converse statement.

%Still, one be careful with Observation \ref{obs_connection}. The singularity arising from the coefficients $a_n$ might not be the right-most singularity, as is implicit from the assumptions of \Propref{prop_resurgence_2}.

%In addition, we would also need to verify that in fact $f$ is recovered from the Laplace transform of $B$, for example by verifying that the assumptions in part \ref{it_connection_nevanlinna(a)} of Nevanlinna's theorem are satisfied.

%All of these problems are avoided by imposing assumptions over $B$ directly and defining $f$ as the Laplace transform of $B$.

\Propref{prop_resurgence_2} is a formalisation of the idea behind \cite{Peris}, where quark-hadron duality was discussed in the framework of Borel transforms. The idea is closely related to alien calculus and, in this context, it was already discussed in \cite[p.~100--101]{Sauzin}.

\begin{figure}
\centering
\subfloat[$f_\theta$ \label{fig_connection_def_ftheta}]{
\begin{tikzpicture}[scale=1]
\fill[fill=gray!20!white] (2,0.25) -- (0,0.25) arc(90:270:0.25) -- (2,-0.25);
\fill[fill=gray!20!white] (0.75,0.5) -- (0.5,2) arc(99.4623:-99.4623:2) -- (0.75,0.5);

\fill (1.25,0.5) circle[radius=2pt] node[right]{$S$};
\draw[dashed] (1.25,0.5) -- (-2,0.5);

\draw[->] (1.75,0) arc (0:44:0.5) node[midway,right]{$\theta$};

\draw[help lines,->] (-2,0) -- (2.5,0) coordinate (xaxis);
\draw[help lines,->] (0,-2) -- (0,2) coordinate (yaxis);

\path[draw,line width=0.8pt,decoration={markings, mark=at position 0.2 with {\arrow{>}}, mark=at position 0.75 with {\arrow{>}}},
postaction=decorate] (0,0) -- (1.25,0) -- (2.66421,1.41421) node[above]{$|B(\zeta)|\le K e^{A|\zeta|}$};

\node[above] at (yaxis) {$\zeta$ plane};
\end{tikzpicture}
}
\hspace{1cm}
\subfloat[$\disc{f}$ \label{fig_connection_disc_f}]{
\begin{tikzpicture}[scale=1]
\fill[fill=gray!20!white] (2,0.25) -- (0,0.25) arc(90:270:0.25) -- (2,-0.25);
\fill[fill=gray!20!white] (0.75,0.5) -- (0.5,2) arc(99.4623:-99.4623:2) -- (0.75,0.5);

\path[draw,line width=0.8pt,decoration={markings, mark=at position 0.75cm with {\arrow{>}},
markings, mark=at position 3.15cm with {\arrow{>}},
mark=at position 4.9cm with {\arrow{>}}},
postaction=decorate] (0.6,2) -- node[midway,right]{$\mathcal{C}_+$} (0.874123,0.636808) arc(160:-160:0.4) node[midway,right]{$C_\delta$} -- node[midway,right]{$\mathcal{C}_-$} (0.6,-2);

\fill (1.25,0.5) circle[radius=2pt] node[right]{$S$};
\draw[dashed] (1.25,0.5) -- (-2,0.5);

\draw[help lines,->] (-2,0) -- (2.5,0) coordinate (xaxis);
\draw[help lines,->] (0,-2) -- (0,2) coordinate (yaxis);

\node[above] at (yaxis) {$\zeta$ plane};
\end{tikzpicture}
}
\caption{Contours of integration in the $\zeta$ plane needed to define $f_\theta$ and $\disc{f}$, respectively, for the proof of \Propref{prop_resurgence_2}. The grey region (minus $S$ and its cut) is the domain where $B$ is analytic.}
\label{fig_connection_f+discf}
\end{figure}
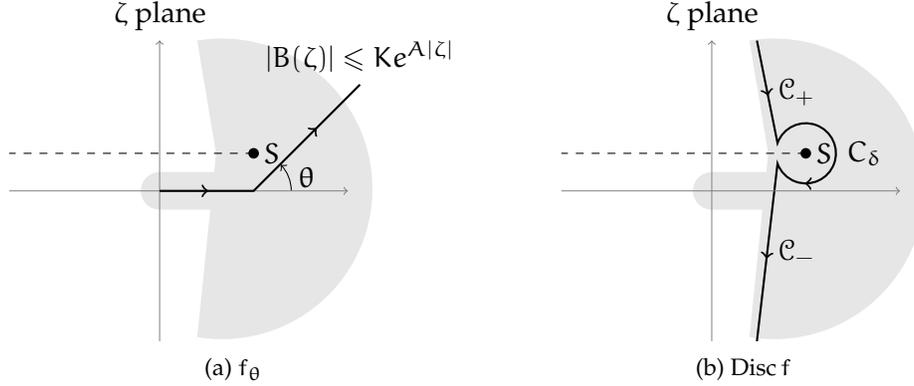

\begin{prop}
\label{prop_resurgence_2}
Given $S\in \mathbb{C}\setminus \mathbb{R}^+$ and $\epsilon>0$, let $B$ be an analytic function in a domain containing $\mathbb{R}^+$ and the sector $|\arg(\zeta-S)|\le \pi/2 + \epsilon$, from which we subtract the point $S$ and the cut arising from the singularity at $S$ (see the grey region in \Figref{fig_connection_f+discf}). Further assume that $|B(\zeta)|\le K e^{A|\zeta|}$ in the above domain and that
\begin{equation}
B(\zeta) = -\frac{b_0}{\pi} \frac{\Gamma(\alpha)}{(\zeta - S)^\alpha}\bigg[ 1 + \mathcal{O}(\zeta - S) \bigg], \quad \zeta \rightarrow S\,.
\label{eq_connection_B_prop_2}
\end{equation}
Then the function
\begin{equation}
f(z) = \int_0^\infty\dd\zeta\, e^{-z\zeta}\, B(\zeta)\,, \quad \Re(z)>A\,,
\label{eq_connection_prop_2_laplace}
\end{equation}
admits two analytic continuations (clockwise and anti-clockwise) around a disc of radius $A$ centred in the origin and their difference yields
\begin{equation}
\disc{f}(z) = 2i\, b_0\, e^{-Sz} (-z)^{\alpha-1}\left[ 1 + \mathcal{O}\left(\frac{1}{z}\right) \right], \quad z\rightarrow -\infty\,.
\label{eq_connection_disc_prop_2}
\end{equation}
\end{prop}

The proof can be found in \Appref{sec_connection_proof_2}. In summary, if $f$ is the Laplace transform in \Eqref{eq_connection_prop_2_laplace}, then we may compute analytic continuations of $f$ by changing the direction of integration. In addition, $\disc{f}$ may be computed from the difference of the Laplace transform between the two directions $-\pi/2-\epsilon$ and $+\pi/2+\epsilon$, as we saw in the second part of \Exref{example_connection_resurgence}. If necessary, we avoid the singularity at $S$ by considering the paths defined in \Figref{fig_connection_def_ftheta}.

Furthermore, the difference between the two Laplace transforms, and therefore $\disc{f}$, can be rewritten as an integral along a single path, as that depicted in \Figref{fig_connection_disc_f}. Let us call $\mathcal{C}$ to this path.

We notice that, given any segment $[a,b]$, we have the bound
\begin{equation}
\left| \int_a^b \dd\zeta\, e^{-z\zeta}\, B(\zeta) \right| \le \int_a^b |\dd\zeta|\, e^{-z\Re(\zeta)}\, |B(\zeta)| \le m\, |b-a|\, e^{-z\max\big\{\Re(a),\,\Re(b)\big\}},
\end{equation}
valid for $z<0$, where $m$ is the maximum of $B$ inside $[a,b]$. This means that, as long as we can deform the path $\mathcal{C}$ to the left of the Borel plane, we can arbitrarily reduce the values of $\Re(a)$ and $\Re(b)$ and, therefore, the power of the exponential behaviour. The rightmost singularity of $B$ (the point $S$) prevents further deformation to the left. Thus, we expect that the singularity at $S$ encodes the leading contribution to $\disc{f}(z)$ for large $-z$.

%As an illustration, in \Figref{fig_connection_ftheta(b)} we could deform $\mathcal{C}$ to the left up to the singularity at $\zeta=S=-1$, thus $\disc{f}(z)$ behaves like $e^{-Sz} = e^{z}$.

It may seem from \Exref{example_connection_resurgence} that we need to know the properties of $B$ (analyticity, bounds) along the contour $C_R$ in \Figref{fig_connection_ftheta(b)}. This is not necessary, as we can confirm from the proof in \Appref{sec_connection_proof_2}. $B$ might not even admit an analytical continuation beyond its original domain of analyticity.

%Because we have demanded $\epsilon>0$, $B$ cannot have a wall of singularities accumulating at $\infty$ with constant real part. In principle there should be no problem with this situation (and we may admit $\epsilon=0$), but all the singularities would then yield contributions of the same order into \Eqref{eq_connection_disc_prop_2}. The leading behaviour would be given by the sum of all those infinite number of contributions.

As long as the Laplace transform in \Eqref{eq_connection_prop_2_laplace} is well defined, we may admit that $B$ has a singularity at $\zeta=0$. This is specially important such as when $f$ is a correlator in quantum chromodynamics, where $z=q^2$ is the momentum of the correlator. In this case, a singularity at $\zeta=0$ is expected (otherwise, the asymptotic expansion of $f$ would be a simple power-expansion in $1/z$, but we know that the structure of the operator product expansion is much richer, since it contains logarithms). In particular, a singularity at $\zeta=0$ makes a non-exponential contribution to the discontinuity of the correlator.

%As long as the singularities of $f$ in the negative real axis of the upper (lower) Riemann sheet have a bounded modulus, the half-planes of convergence of the Laplace transform can go around the furthest singularity. Thus, $B$ cannot have a wall of singularities in $\Im(\zeta)<0$ ($\Im(\zeta)>0$).

%In other words, if $B$ has a wall of singularities, then $f$ has singularities accumulating at $\infty$ along the negative real axis (in at least one of the sheets). The converse is also true, as we have seen in the proof of \Propref{prop_resurgence_2}.

%For example, the derivative of the digamma function was discussed in \cite{Peris}. The Borel transform of this function has a wall of singularities: $\zeta = 2\pi i k$, with $k\in \mathbb{Z}\setminus \{0 \}$. This has to be so, because the derivative of the digamma function has poles at all negative integer positions.

%%%%%%%%%%%%%%%%%%%%%%%%%%%%%%%%%%%%%%%%%%%%%%%%%%%%%%%%%%%%%%%%%%%%%%%%%%%%%%%%%%%%%%%%%%%%%%%%%%%%

\section{Singularities on the positive real axis of the Borel plane}
\label{sec_connection_median}
Until now, culminating in \Propref{prop_resurgence_1}, we have defined the resurgent connection as the relation between the leading behaviours of $\disc{f}$ and $a_n$. Even if $f$ satisfies the bound $|f(z)|\le |a_0|/|z|$ (or a dispersion relation), the resurgent connection cannot take place in the situation of Observation \ref{obs_connection}. However, this situation arises because our knowledge of $\disc{f}$ is not complete. If we knew the exact discontinuity, we would be able to determine  all the singularities in the Borel plane and, therefore, determine the large order behaviour of the $a_n$ to any desired accuracy. In this case, Observation \ref{obs_connection} becomes meaningless.

With this in mind, from the theoretical point of view we find it more convenient to define resurgence in the following way: \guillemotleft $f$ satisfies a resurgent connection if $\disc{f}$ fully encodes all the information about the coefficients $a_n$\guillemotright{}. Then, the question whether $f$ satisfies a resurgent connection reduces to the question whether the asymptotic expansion of $f$ is missing exponential corrections or not. In this sense, the discussion is simplified, as we no longer have to deal with Observation \ref{obs_connection}. For instance, the function in \Appref{sec_connection_ordering} would have a resurgent connection, because even if the singularities in its Borel transform, $S_1$ and $S_2$, are such that $\Re(S_1) < \Re(S_2)$, but $|S_1|<|S_2|$, the exact coefficients $a_n$ of its asymptotic expansion are in correspondence with its exact discontinuity.

The main lesson from \Secref{sec_connection_formal} is then that the resurgent connection (as defined in the previous paragraph) is naturally satisfied by functions expressible as a Laplace transform. A Laplace transform satisfies by default the assumptions in part \ref{it_connection_nevanlinna(a)} of Nevanlinna's theorem and, as such, exponential corrections hidden beyond its asymptotic expansion are forbidden. From this observation, it is natural that Laplace transforms always feature a resurgent connection, because the absence of exponential corrections also ensures that these cannot incorporate additional discontinuities that may spoil the connection (as we saw happening in \Exref{example_connection_exp_correction}).

Nevertheless, if the exponential corrections do not incorporate discontinuities (for example, because they are entire functions), then it is clear that the resurgent connection will take place even for functions which are not expressible as a Laplace transform.

In the present section, we will exemplify that median resummed series, which by definition incorporate exponential corrections, still satisfy the resurgent connection. Median resummed series are introduced in \cite{AnicetoSchiappa, Unsal}. They arise from the necessity to assign finite and purely real values (when $z>0$) to divergent series whose Borel transforms have singularities on the positive real axis.

\begin{example}
\label{example_connection_median_resummation}
In this example we want to define the median resummation of $\sum n!/z^{n+1}$ and discuss its resurgent connection. We have seen that the Borel transform of the above series is $B(\zeta) = 1/(\zeta-1)$. The Laplace transform of $B$ along the direction $\theta=\pi$ defines a function in $\Re(z)<0$ which can be analytically extended to $\Re(z)>0$. Due to the pole at $\zeta=1$, this function takes values with non-zero imaginary part for $z>0$ and, also, this imaginary part is ambiguous depending on the path of analytic continuation.

The median resummation of $\sum n!/z^{n+1}$ is defined by
\begin{equation}
f(z) = \int_0^{\infty e^{i\theta}}\dd\zeta\, e^{-z\zeta}\, B(\zeta) \pm i\pi\, e^{-z}\,,
\label{eq_connection_median_sum}
\end{equation}
where the minus sign is chosen when $\theta \in (0,+\pi)$ and the plus sign, when $\theta\in (-\pi,0)$.\footnote{It might be argued that, for $\theta \in (+\pi/2,+\pi) \cup (-\pi,-\pi/2)$, there should be no exponential correction because there are no singularities in $\mathbb{R}^-$ and, thus, no ambiguity to cancel. Removing the exponential term for those values of $\theta$ would break the analytical properties of the resummed series $f(z)$.} By construction, $f$ is non-ambiguous and purely real for $z>0$.

We denote by $f_0$ the Laplace transform in \Eqref{eq_connection_median_sum} (that is, without the exponential term). On one hand, $f_0$ alone satisfies the assumptions of \Propref{prop_resurgence_1} (actually $f_0(-z)$, but we can change the variable again after determining the connection). But on the other hand, the exponential term spoils the condition that $|f(z)| \le |a_0|/|z|$ when $z$ goes along the imaginary axis.

By writing $f_0$ in terms of the exponential integral $E_1$, we obtain
\begin{equation}
f(z) = -e^{-z}E_1(-z) \pm i\pi\, e^{-z} = -e^{-z}\left( -\gamma - \log(z) - \sum_{k\ge 1} \frac{z^k}{k!\, k} \right).
\label{eq_connection_ambiguity_cancellation}
\end{equation}
In the second equality, we have used \Eqref{eq_connection_E1} and absorbed the ambiguous exponential term inside the logarithm. This process changes the logarithm from $\log(-z)$ to $\log(z)$.

\Eqref{eq_connection_ambiguity_cancellation} explicitly shows that the ambiguous exponential term that arises from median resummation does not alter the discontinuity of the Laplace transform $f_0$, whose multivalued component is $\log(-z)$. Indeed, $\log(-z)$ and $\log(z)$, albeit being different functions, have the same discontinuity.
\end{example}

We gather the following observations from \Exref{example_connection_median_resummation}:
\begin{enumerate}[label=(\Alph*)]
\item \label{it_connection_property_(A)} $f_0$ satisfies the resurgent connection in \Propref{prop_resurgence_1}.
\item \label{it_connection_property_(B)} For $z\rightarrow +\infty$, $f_0$ shares the same asymptotic expansion with $f$.
\item \label{it_connection_property_(C)} $\disc{f_0}(z) = \disc{f}(z)$ for $z\in\mathbb{C}$ (in particular $z<0$) by choosing appropriate branches for $\log(z)$ and $\log(-z)$.
\end{enumerate}
We conclude from the above points that $f$ satisfies the same resurgent connection as the Laplace transform $f_0$.

To finish this section, we will exemplify that we expect a similar situation for path integrals with singularities on the positive axis of the Borel plane. We consider a field toy model in which space-time is 0-dimensional and fields are real. That is, fields are functions with domain equal to a single point $\{ p \}$ and we can identify each field configuration $\phi(p) = \phi$ as the number $\phi \in \mathbb{R}$ that the field configuration takes at $p$. We further specify the theory by setting the action
\begin{equation}
S(\phi,\lambda) =
\begin{dcases}
\frac{1}{2\lambda}\sin^2\big(\sqrt{\lambda}\phi\big) & \text{if } \phi \in \left[-\frac{\pi}{2\sqrt{\lambda}}, +\frac{\pi}{2\sqrt{\lambda}}\right]\\[1mm]
0 & \text{otherwise.}
\end{dcases}
\end{equation}

In the path integral approach, integration along all field configurations is an integration along $\mathbb{R}$. The partition function of this model is
\begin{equation}
Z(\lambda) = \frac{1}{\sqrt{2\pi}}\int_{-\frac{\pi}{2\sqrt{\lambda}}}^{+\frac{\pi}{2\sqrt{\lambda}}} e^{-S(\phi,\lambda)}\, \dd\phi = \sqrt{\frac{\pi}{2}}\frac{ e^{-\tfrac{1}{4\lambda}}}{\sqrt{\lambda}} I_0\Big( \frac{1}{4\lambda} \Big)\,,
\label{eq_connection_sine_partition}
\end{equation}
where $I_n$ is the modified Bessel function of the first kind, which is an entire function.

We will compute the perturbative and the non-perturbative expansions of the path integral $Z$. The perturbative part comes from quantum fluctuations around the trivial saddle point (zero action), while the non-perturbative part comes from fluctuations around the non-trivial saddle point (non-zero action).

Extrema coincide with saddle points in the case of analytic functions. The points $\phi$ that extremise the action are given by
\begin{equation}
\frac{\dd S(\phi,\lambda)}{\dd\phi} = \frac{\sin\big( \sqrt{\lambda}\phi \big)\cos\big( \sqrt{\lambda}\phi \big)}{\sqrt{\lambda}} = 0\,,
\end{equation}
and this equation has the solutions
\begin{equation}
\phi_0 = 0\,,\quad\phi_1 = \,\pm \frac{\pi}{2\sqrt{\lambda}}\,.
\end{equation}

On one hand, quantum fluctuations around the trivial saddle point $\phi_0$ yields \cite{Unsal}
\begin{equation}
Z_{\textrm{P}}(\lambda) = \frac{1}{\lambda}\sum_{n\ge 0}a_n \lambda^{n+1}\,,
\label{eq_connection_sine_expansion_P}
\end{equation}
with coefficients
\begin{equation}
a_n = \frac{\Gamma\big( n + \frac{1}{2} \big)^2}{\Gamma\big( \frac{1}{2} \big)^2 n!}\,2^n\,.
\label{eq_connection_sine_coefficients}
\end{equation}
On the other hand, quantum fluctuations around the non-trivial saddle point $\phi_1$ (with either of the signs) yields the non-perturbative expansion \cite{Unsal}
\begin{equation}
Z_\textrm{NP}(\lambda) = 2 i e^{-\tfrac{1}{2\lambda}} \frac{1}{\lambda} \sum_{n\ge 0}b_n \lambda^{n+1}\,,
\label{eq_connection_sine_expansion_NP}
\end{equation}
where $b_n = (-1)^n a_n$. Here we have summed the asymptotic expansions from both saddle points $\phi_1$, thus the factor 2 arises.

To continue the discussion, we define
\begin{equation}
f(z) = \frac{Z(1/z)}{z}\,,
\end{equation}
such that $z=1/\lambda$ and the asymptotic expansion of $f$ starts with the power $1/z$.

From the last expression in \Eqref{eq_connection_sine_partition}, we see that $f$ has a square root type singularity at $z=0$ and that
\begin{equation}
\disc{f}(z) = 2f(z)\,, \quad z\in \mathbb{C}\,.
\label{eq_connection_sine_Z_monodromy}
\end{equation}
Furthermore, one can check the validity of the following asymptotic expansions:
\begin{equation}
\begin{array}{cc}
\displaystyle f(z) \sim \sum_{n\ge 0}\frac{a_n}{z^{n+1}}\,, \quad \Re(z)>0\,; & \displaystyle f(z) \sim \pm i e^{-\tfrac{z}{2}} \sum_{n\ge 0}\frac{b_n}{z^{n+1}}\,, \quad \Re(z)<0\,.
\end{array}
\label{eq_connection_transseries_split}
\end{equation}
The $\pm$ encodes the branch cut of $f$. We take the plus sign for $z$ above the cut and the minus sign for $z$ below the cut.

The above result can be compactly written as the transseries
\begin{equation}
f(z) \sim \sum_{n\ge 0}\frac{a_n}{z^{n+1}} \pm i e^{-\tfrac{z}{2}} \sum_{n\ge 0}\frac{b_n}{z^{n+1}}\,, \quad |z|\rightarrow \infty\,.
\label{eq_connection_transseries}
\end{equation}
The transseries contains both the perturbative expansion $Z_\textrm{P}$ and the non-perturbative corrections $Z_\textrm{NP}$. When $\Re(z)>0$, the second term is exponentially suppressed and it is hidden for large $|z|$. Similarly, for $\Re(z)<0$, the second term is exponentially enhanced and the first term is now hidden.

The exponential corrections in \Eqref{eq_connection_transseries} would in general spoil the resurgent connection. However, we still verify that the large order behaviour of \Eqref{eq_connection_sine_coefficients}
\begin{equation}
a_n \sim \frac{\Gamma(n)}{\pi}\, 2^n\,, \quad n\rightarrow \infty\,,
\label{eq_connection_sine_an_beh}
\end{equation}
is in correspondence with
\begin{equation}
\disc{f}(z) = 2f(z) \sim 2i \frac{e^{-\tfrac{z}{2}}}{z}\,, \quad z\rightarrow -\infty\,.
\label{eq_connection_phi_disc}
\end{equation}
(With parameters $b_0 = 1$, $\alpha=0$, $S=1/2$ in \Propref{prop_resurgence_1}).\footnote{We only display the connection at leading order, but it is easy to check that the connection also happens between all sub-leading corrections to \Eqref{eq_connection_sine_an_beh} and \Eqref{eq_connection_phi_disc}.}

%Later we will understand why the resurgence connection is still realized in this example. Right now, we make the following observation: the coefficients $a_n$ are connected with the exponential corrections inside the transseries of \Eqref{eq_connection_transseries}. The determining factor for this to be true is that $\disc{f}$ is proportional to $f$ itself, thus the resurgence connection between these two functions is actually a connection between complementary $z$ regimes ($\Re(z)>0$ and $\Re(z)<0$) of $f$.

\begin{sloppypar}
Let us argue why this resurgent connection takes place. From the transseries in \Eqref{eq_connection_transseries}, we may write $f$ as the median resummation of the series $\sum a_n/z^{n+1}$:
\begin{equation}
f(z) = \int_0^{\infty e^{i\theta}}\! \dd\zeta\, e^{-z\zeta}\, B_0(\zeta) \pm i e^{-\tfrac{z}{2}} \int_0^{\infty e^{i\theta}}\! \dd\zeta\, e^{-z\zeta}\, B_1(\zeta)\,,
\label{eq_connection_sum_transseries}
\end{equation}
where
\begin{equation}
\begin{array}{ll}
B_0(\zeta) = F\left( \tfrac{1}{2},\tfrac{1}{2}, 1, 2\zeta \right)\,, & B_1(\zeta) = B_0(-\zeta)\,,
\end{array}
\label{eq_connection_sine_borel}
\end{equation}
are the Borel transforms of the two asymptotic expansions in \Eqref{eq_connection_transseries_split} ($F$ being the ordinary hypergeometric function). The minus sign is chosen when $\theta \in (0,+\pi)$, and the plus sign, when $\theta \in (-\pi,0)$.
\end{sloppypar}

Let us call $f_0$ the Laplace transform of $B_0$ as it appears in \Eqref{eq_connection_sum_transseries}. We note that $B_0(\zeta)$ has a logarithmic singularity at $\zeta=1/2$. This singularity generates an ambiguous imaginary part in $f_0(z)$ for $z>0$, which is cancelled against the second term in \Eqref{eq_connection_sum_transseries}. In this way, $f(z)$ is real for $z>0$. This is known as ambiguity cancellation of the Borel sum.

In this example we will prove that, instead of \ref{it_connection_property_(C)}, we have
\begin{enumerate}[label=(\Alph*')]
\setcounter{enumi}{2}
\item \label{it_connection_property_(C')}$\disc{f_0}(z)$ and $\disc{f}(z)$ (understood as analytic functions in the variable $z$) differ by an exponentially small term when $z\rightarrow -\infty$. Equivalently, the two discontinuities share the same asymptotic expansion in that limit.
\end{enumerate}
Gathering \ref{it_connection_property_(A)}, \ref{it_connection_property_(B)}, \ref{it_connection_property_(C')}, the conclusion of \Exref{example_connection_median_resummation} follows again: $f$ satisfies the same resurgent connection as the Laplace transform $f_0$. So, let us prove \ref{it_connection_property_(C')}. 

The discontinuity of $f_0$ comes from the difference in the two directions above ($\theta_+$) and below ($\theta_-$) the positive real axis of the Borel plane:
\begin{equation}
\begin{aligned}
\disc{f_0}(z) &= \left( \int_0^{\infty e^{i\theta_\plus}} - \int_0^{\infty e^{i\theta_\minus}}\right) \dd\zeta\, e^{-z\zeta}\, B_0(\zeta)\\
&= -e^{-\tfrac{z}{2}} \int_0^\infty \dd\zeta\, e^{-z\zeta}\, \disc{B_0}\left(\zeta+ \tfrac{1}{2}\right)\,, \quad \Re(z)>0\,.
\end{aligned}
\label{eq_connection_disc_f0_1}
\end{equation}
The second term in \Eqref{eq_connection_sum_transseries} exactly cancels this discontinuity in $\Re(z)>0$. The $\pm$ conspire with the singularity in $B_0$ so that when changing the direction $\theta$ from below 0 to above 0, there is effectively no singularity in $B_0$. Specifically, we have the cancellation
\begin{equation}
\disc{f_0}(z) - 2ie^{-\tfrac{z}{2}}\int_0^\infty \dd\zeta\, e^{-z\zeta}\, B_1(\zeta) = 0\,, \quad \Re(z)>0\,.
\label{eq_connection_cancellation}
\end{equation}
The fulfilment of this equation in $\Re(z)>0$ implies that
\begin{equation}
\disc{B_0}\left(\zeta+ \tfrac{1}{2}\right) = -2i B_1(\zeta)\,, \quad \zeta \ge 0\,,
\label{eq_connection_B0_B1}
\end{equation}
and the result also has to be satisfied for $\zeta$ in the Riemann surface of $B_1$ due to the unicity of the analytic continuation.

Combining \Eqref{eq_connection_disc_f0_1} and \Eqref{eq_connection_B0_B1}, we may write
\begin{equation}
\disc{f_0}(z) = 2 i e^{-\tfrac{z}{2}} \int_0^{\infty e^{i\theta}}\! \dd\zeta\, e^{-z\zeta}\, B_1(\zeta) \,, \quad \Re\big(z e^{i\theta}\big)>0\,.
\label{eq_connection_disc_f0_2}
\end{equation}
We also introduced a direction $\theta$ in the integral such that $\disc{f_0}$ can be analytically continued from $\Re(z)>0$ to $\Re(z)<0$.

Thanks to the ambiguity cancellation, \Eqref{eq_connection_sum_transseries} changes smoothly when $\theta$ changes from above to below the positive real axis (even if the singularity in $\zeta=1/2$ of $B_0$ would normally prevent that). Instead, the directions above ($\theta_+$) and below ($\theta_-$) the negative real yield two different results for \Eqref{eq_connection_sum_transseries} at the same point $z$. Their difference is defined as the discontinuity of $f$. Namely, we have
\begin{equation}
\disc{f}(z) = i e^{-\tfrac{z}{2}}\left( \int_0^{\infty e^{i\theta_\minus}} + \int_0^{\infty e^{i\theta_\plus}}\right) \dd\zeta\, e^{-z\zeta}\, B_1(\zeta)\,, \quad \Re(z)<0\,.
\label{eq_connection_f_disc_resurgence}
\end{equation}
Note that the two integrals would normally appear with opposite sings, but because of the $\pm$, the integrals are summed instead.

Now there are two ways to proceed. We either check that $\disc{f_0}$ and $\disc{f}$ differ by an exponentially small term or that they share the same asymptotic expansion. Here we will go with the later. Using part \ref{it_connection_nevanlinna(b)} of Nevanlinna's theorem in the Laplace transform appearing either in \Eqref{eq_connection_disc_f0_2} (with $\theta$ close enough to $\pi$) or in \Eqref{eq_connection_f_disc_resurgence}, we see that in both cases
\begin{equation}
\begin{rcases}
\disc{f_0}(z)\\
\disc{f}(z)
\end{rcases}
\sim 2i e^{-\tfrac{z}{2}}\sum_{n\ge 0} \frac{b_n}{z^{n+1}}\,, \quad z\rightarrow -\infty\,,
\end{equation}
where $b_n = B_1^{(n)}(0)$. This completes the verification.\footnote{It is interesting to note that $\disc{f_0} \neq \disc{f}$. In fact, the two discontinuities differ by a non-zero exponential correction that arises from the singularity at $\zeta=-1/2$ in $B_1$.}

It might be argued that the cancellation in \Eqref{eq_connection_cancellation} is a particular feature of the example we have discussed, but in fact, this is a general feature of median resummed series, which ensures that the resulting function is real for $z>0$.

We also may make the following observation from this example. Even if the resurgent connection is originally a connection between the asymptotic expansion of a function and its discontinuity, in this case it is reinterpreted as a connection between the asymptotic expansion of the function and its exponential corrections.

We finish this section with an important observation. Given an arbitrary series, we haven seen that the function defined as the median resummed series always features a resurgent connection. Nevertheless, given the asymptotic expansion of a function, it could be misleading to think that the median resummation of this expansion always yields the true function. If the median resummed expansion does not coincide with the function, we cannot make any claim regarding its resurgent connection.

In \Appref{sec_connection_resurgence_fails}, we discuss an example within 2-dimensional field theory where the function under study does not coincide with the median resummation of its asymptotic expansion and, in consequence, the function does not satisfy any resurgent connection.

%%%%%%%%%%%%%%%%%%%%%%%%%%%%%%%%%%%%%%%%%%%%%%%%%%%%%%%%%%%%%%%%%%%%%%%%%%%%%%%%%%%%%%%%%%%%%%%%%%%

\section{Conclusions}
In the present work, resurgence is defined as a connection between the discontinuity of a function and the coefficients of its asymptotic expansion (\Propref{prop_resurgence_1}). These two elements are related through the singularities in the Borel plane. Schematically:
\begin{center}
\begin{tabular}{ccccc}
Discontinuity of $f$
&
$\leftrightharpoons$
&
Singularities of $B$
&
$\leftrightharpoons$
&
\makecell{Coefficients $a_n$ in the \\
asymptotic expansion of $f$}
\end{tabular}
\end{center}
In the literature (for instance, \cite{DunneUnsal}), resurgence is understood as a connection between different exponential sectors of a transseries. Discussing the 0-dimensional path integral of \Eqref{eq_connection_sine_partition}, we have concluded that these are two sides of the same coin. At this point, this might seem a trivial statement, because in that example the exponential corrections in the transseries encode the asymptotic behaviour of the discontinuity (compare \Eqref{eq_connection_transseries} to \Eqref{eq_connection_phi_disc}).

To reach the above conclusion, we had to make a detour. We first discussed sufficient conditions that forbid the existence of exponential corrections to the asymptotic expansion of a function, a result known as Nevanlinna's theorem. Under the assumptions of the theorem, a function can be uniquely recovered from its asymptotic expansion with the method of Borel summation. This theorem was interesting in the context of our work, because we later showed that a function free of exponential corrections (thus expressible as a Laplace transform) is the minimal unit featuring a resurgent connection.

We later observed that we can add exponential corrections to these minimal units as long as the corrections do not spoil the original resurgent connection by incorporating additional discontinuities (a canonical example is given when the exponential corrections are entire functions). In particular, we have discussed a special case of exponential corrections which arise from the median resummation of a divergent series with fixed sign coefficients. For all practical purposes, the resurgent connection held in our examples of median resummation as if the exponential corrections were not even present.

%We have seen that we cannot in general recover a function only from its asymptotic expansion, because hidden exponential corrections incorporate information which is essential to the reconstruction of the function. While the asymptotic expansion and the exponential corrections are interconnected for median resummed series, they are in general decoupled given an arbitrary function.

Finally, we want to remark again that, in general, a function defined by a path integral does not have to coincide with the median resummation of its asymptotic expansion. For example, in the 2-dimensional path integral of \Appref{sec_connection_resurgence_fails}, it is clear that the median resummation of the asymptotic expansion does not recover the full function. It is beyond the scope of our work to understand when median resummation is enough to recover the true function.

%%%%%%%%%%%%%%%%%%%%%%%%%%%%%%%%%%%%%%%%%%%%%%%%%%%%%%%%%%%%%%%%%%%%%%%%%%%%%%%%

\section*{Acknowledgements}
I am grateful to Santiago Peris for fruitful discussions and for carefully reading this work during each stage of its writing. I also want to thank Matthias Jamin for his comments.  My work has been supported by Ministerio de Economía y Competitividad Grant number CICYT-FEDER-FPA2014-55613-P and CICYT-FEDER-FPA2017-86989-P.

%%%%%%%%%%%%%%%%%%%%%%%%%%%%%%%%%%%%%%%%%%%%%%%%%%%%%%%%%%%%%%%%%%%%%%%%%%%%%%%%%%%%

\appendix
\section{An illustration of Observation \ref{obs_connection}}
\label{sec_connection_ordering}
We define
\begin{equation}
f(z) = -e^{-S_1 z} E_1(-S_1 z) - e^{-S_2 z} E_1(-S_2 z)\,.
\end{equation}
Each component of the function has a branch cut conventionally placed along the direction $-\arg(S_1)$ and $-\arg(S_2)$, respectively. Similarly to what we did in \Eqref{eq_connection_F_cut}, we may place the cut along $\mathbb{R}^-$ by adding appropriate exponential terms. By assuming $\Re(S_1)$, $\Re(S_2)<0$, we ensure that these exponential terms never become enhanced for any $z\in \mathbb{C}$.

We have constructed this function so that its asymptotic expansion is
\begin{equation}
f(z) \sim \sum_{n\ge 0} \frac{a_n}{z^{n+1}}\,,  \quad |z|\rightarrow \infty\,,
\end{equation}
with
\begin{equation}
a_n = n! \left( \frac{1}{S_1^{n+1}} + \frac{1}{S_2^{n+1}} \right).
\label{eq_connection_obs_an}
\end{equation}
This function satisfies a dispersion relation, because it goes like $1/z$ for large $|z|$. Its discontinuity, arising from the logarithmic term in \Eqref{eq_connection_E1}, is given by
\begin{equation}
\disc{f}(z) = 2 \pi i \left( e^{-S_1 z} + e^{-S_2 z} \right).
\label{eq_connection_obs_disc}
\end{equation}

On one hand, the leading behaviour for large $n$ in \Eqref{eq_connection_obs_an} is given by
\begin{equation}
a_n \sim
\begin{dcases}
\frac{n!}{S_1^{n+1}} & \text{if } |S_1|<|S_2|\\
\frac{n!}{S_2^{n+1}} & \text{if } |S_2|<|S_1|\,.
\end{dcases}
\end{equation}
On the other hand, the leading behaviour for large $-z$ in \Eqref{eq_connection_obs_disc} is given by
\begin{equation}
\disc{f}(z) \sim 2\pi i
\begin{dcases}
e^{-S_1 z} & \text{if } \Re(S_1)>\Re(S_2)\\
e^{-S_2 z} & \text{if } \Re(S_2)>\Re(S_1)\,.
\end{dcases}
\end{equation}
Clearly, there is no correspondence between the two leading behaviours if $|S_1|<|S_2|$ and $\Re(S_1)<\Re(S_2)$, or the other way around.

%%%%%%%%%%%%%%%%%%%%%%%%%%%%%%%%%%%%%%%%%%%%%%%%%%%%%%%%%%%%%%%%%%%%%%%%%%%%%%%%%%%%%%%%%%%%%%%%%%%

\section{Proof of \Propref{prop_resurgence_1}}
\label{sec_connection_proof_1}
%Proof 1
First of all, we will check that the function
\begin{equation}
B(\zeta) = \frac{1}{2\pi i}\int_{\mathcal{C}_a}\! \dd z\, e^{z\zeta}\, f(z)\,,
\label{eq_connection_B_def_proof1}
\end{equation}
with $a>A$, initially defined for $\zeta>0$, can be extended to an analytic function in $\Re(\zeta)>S$.

Using $|f(z)| \le |a_0|/|z|$, we can deform $\mathcal{C}_a$ to a contour surrounding the cut of $f$, similar to \Figref{fig_connection_dispersion}, but with the origin translated to the point $A$. We obtain
\begin{equation}
B(\zeta) = \frac{1}{2\pi i}\int_{A-\delta}^{-\infty}\! \dd z\, e^{z\zeta}\, \disc{f}(z) + \frac{1}{2\pi i}\int_{C_\delta(A)} \! \dd z\, e^{z\zeta}\, f(z)\,,
\label{eq_connection_B_def2_proof1}
\end{equation}
where $C_\delta(A)$ is a circle of radius $\delta$ around $A$. We denote the second term by $E(\zeta)$. It is entire in $\zeta$, because $C_\delta(A)$ is bounded. Choosing $\delta$ large enough and using \Eqref{eq_connection_disc_prop}, it is easy to check that the first term is absolutely convergent in $\Re(\zeta)>S$, and thus it defines an analytic function there.

In particular, given that $S<0$, $B$ is analytic at $\zeta=0$ and we may compute any number of derivatives at this point. We have
\begin{equation}
\begin{aligned}
B(\zeta) &= \frac{1}{2\pi i}\int_{-\delta}^{-\infty} \! \dd z\sum_{n\ge 0} \frac{(z\zeta)^n}{n!}\, \disc{f}(z) + E(\zeta)\\
&= \sum_{n\ge 0} \frac{\zeta^n}{n!} \left[ \frac{1}{2\pi i}\int_{-\delta}^{-\infty} \dd z \, z^n\, \disc{f}(z)\right] + E(\zeta)\,.
\end{aligned}
\label{eq_connection_B}
\end{equation}
Here it is correct to commute the sum with the integral as a consequence of the dominated convergence theorem. From \Eqref{eq_connection_B}, we may read the $n$-th derivative of $B$ at 0:
\begin{equation}
\begin{aligned}
B^{(n)}(0) &= \frac{1}{2\pi i}\int_{-\delta}^{-\infty} \! \dd z \, z^n\, \disc{f}(z) + E^{(n)}(0)\\
&= \frac{(-1)^{n+1}}{\pi}\, b_0 \int_\delta^\infty \dd z\, e^{Sz} z^{n+\alpha-1}\big[ 1 + R(z)\big] + E^{(n)}(0)\\
&=\begin{multlined}[t][0.75\textwidth]
\frac{(-1)^{n+1}}{\pi}\, b_0 \bigg[ \frac{\Gamma(n+\alpha)}{(-S)^{n+\alpha}} - \frac{\gamma(n+\alpha,-\delta S)}{(-S)^{n+\alpha}}\\
+ \int_\delta^\infty \dd z\, e^{Sz}\, z^{n+\alpha-1} R(z) \bigg] + E^{(n)}(0)\,,
\end{multlined}
\end{aligned}
\label{eq_connection_B(n)}
\end{equation}
where $\gamma(s,x)$ is the lower incomplete gamma function and $|R(z)|\le L/|z|$.

We want to check that $B^{(n)}(0)$ satisfies \Eqref{eq_connection_an_prop}. That is, defining the leading contribution
\begin{equation}
a'_n = \frac{(-1)^{n+1}}{\pi}\frac{\Gamma(n+\alpha)}{(-S)^{n+\alpha}}\, b_0\,,
\end{equation}
we want to prove that
\begin{equation}
\left| \frac{B^{(n)}(0)}{a'_n} - 1  \right| \le \frac{M}{n}
\label{eq_connection_B(n)_bound}
\end{equation}
for large enough $n$. The proof follows from \Eqref{eq_connection_B(n)}. We have
\begin{multline}
\left| \frac{B^{(n)}(0)}{a'_n} - 1  \right| \le \frac{|\gamma(n+\alpha,-\delta S)|}{|\Gamma(n+\alpha)|} + \frac{(-S)^{n+\Re(\alpha)}\left|\int_\delta^\infty \dd z\, e^{Sz}\, z^{n+\alpha-1} R(z)\right|}{|\Gamma(n+\alpha)|}\\
+ \frac{(-S)^{n+\Re(\alpha)}\, \big| E^{(n)}(0) \big|}{|\Gamma(n+\alpha)|}\,.
\end{multline}

We consider the lower bound $|\Gamma(n+\alpha)| \ge M_0 \Gamma(n+\Re(\alpha))$, valid for small enough $M_0$ and large enough $n$.
\begin{itemize}
\item Using the asymptotic behaviour $\gamma(s,x) \sim x^s e^{-x}/s$, valid for large $s$, the first term is bounded by
\begin{equation}
\frac{|\gamma(n+\alpha,-\delta S)|}{|\Gamma(n+\alpha)|} \le \frac{M_1(-\delta S)^n}{\Gamma(n+\Re(\alpha)+1)} = \mathcal{O}\left(\frac{1}{n}\right).
\label{eq_connsection_bound_1}
\end{equation}

\item Using $|R(z)|\le L/|z|$, the second term is bounded by
\begin{multline}
\frac{(-S)^{n+\Re(\alpha)}\left|\int_\delta^\infty \dd z\, e^{Sz}\, z^{n+\alpha-1} R(z)\right|}{|\Gamma(n+\alpha)|}\\
\le \frac{M_2(-S)^{n+\Re(\alpha)} \int_0^\infty \dd z\, e^{Sz}\, z^{n+\Re(\alpha)-2}}{\Gamma(n+\Re(\alpha))} =
\mathcal{O}\left(\frac{1}{n}\right).
\label{eq_connsection_bound_2}
\end{multline}

\item Cauchy inequality yields $\big| E^{(n)}(0) \big| \le m\, n!/r^n$, where $m$ is the maximum of $E(\zeta)$ for $\zeta$ along a circle of centre 0 and radius $r$ contained inside the region of analitycity of $E$. Because $E$ is entire, we are free to choose any $r$, in particular we may choose $r=-S+1>0$. Then the third term is bounded by
\begin{equation}
\frac{(-S)^{n+\Re(\alpha)}\, |E^{(n)}(0)|}{|\Gamma(n+\alpha)|} \le \frac{ M_3 \Gamma(n+1)}{\Gamma(n+\Re(\alpha))} \left(\frac{-S}{-S+1}\right)^n = \mathcal{O}\left(\frac{1}{n}\right).
\label{eq_connsection_bound_3}
\end{equation}
\end{itemize}

From the above results, \Eqref{eq_connection_B(n)_bound} is realised. Notice that the first and third bounds can  be improved to an arbitrary power of $1/n$ (the first is like $1/n!$ and the second, like $1/R^n$, $R>1$). Actually only the second bound gives the error in \Eqref{eq_connection_B(n)_bound}.

It is also in the second bound where the hypothesis $S<0$ is needed. Otherwise, the remainder term could yield contributions of higher order than $a_n'$. This is related to Observation \ref{obs_connection}.

\begin{sloppypar}
To complete the proof, we still have to check that $|B(\zeta)| \le K e^{A\Re(\zeta)}$ in $\Re(\zeta)>S$ and that, for any $|\theta| < \pi/2$,
\begin{equation}
f(z) = \int_0^{\infty e^{i\theta}}\! \dd\zeta\, e^{-z\zeta}\, B(\zeta)\,,\quad \Re\big( ze^{i\theta} \big)> a\cos(\theta)\,.
\label{eq_connection_f_theta_prop_1}
\end{equation}
Then part \ref{it_connection_nevanlinna(b)} of Nevanlinna's theorem gives the asymptotic expansion $f(z) \sim \sum B^{(n)}(0)/z^{n+1}$, uniformly valid in $\Re\big( ze^{i\theta} \big)> a\cos(\theta)$.
\end{sloppypar}

\begin{sloppypar}
First we will prove the exponential bound on $B$. The second term in \Eqref{eq_connection_B_def2_proof1} is bounded by $|E(\zeta)| \le K_1' e^{A\Re(\zeta)}e^{\delta|\zeta|}$ for any $\delta>0$. Choosing $\delta = 1/|\zeta|$, we find $|E(\zeta)| \le K_1 e^{A\Re(\zeta)}$. Furthermore, using \Eqref{eq_connection_disc_prop}, we bound the first term in \Eqref{eq_connection_B_def2_proof1} with $K_2 e^{A\Re(\zeta)}$ (also with the same choice $\delta = 1/|\zeta|$). Thus, $B$ is bounded by
\begin{equation}
|B(\zeta)| \le K e^{A\Re(\zeta)}\,.
\label{eq_connection_B_bound}
\end{equation}
\end{sloppypar}

We choose $\theta \in (-\pi/2,+\pi/2)$ and deform the path $\mathcal{C}_a$ into a path $\mathcal{C}_a(\theta)$ parametrized by $w = a + x e^{i(\pi/2-\theta)}$, with $x \in \mathbb{R}$. Assuming that the order of integration can be interchanged, we have, for $\Re\big( ze^{i\theta} \big)>a\cos(\theta)$,
\begin{equation}
\begin{aligned}
\int_0^{\infty e^{i\theta}} \! \dd\zeta\, e^{-z\zeta}\, B(\zeta) &= \int_0^{\infty e^{i\theta}} \! \dd\zeta\, e^{-z\zeta}\, \left( \frac{1}{2\pi i} \int_{\mathcal{C}_a(\theta)} \! \dd w\, e^{w\zeta}\, f(w)\right)\\
&= \frac{1}{2\pi i} \int_{\mathcal{C}_a(\theta)} \! \dd w\, f(w) \int_0^{\infty e^{i\theta}} \! \dd\zeta\, e^{-(z-w)\zeta}\\
&= \frac{1}{2\pi i} \int_{\mathcal{C}_a(\theta)} \! \dd w\, \frac{f(w)}{z-w}\,.
\end{aligned}
\label{eq_connection_B=f}
\end{equation}
In the last line, we deform the path $\mathcal{C}_a(\theta)$ into a circle around $w=z$. This yields $f(z)$ when using the residue theorem. Here we needed that $|f(z)|\le |a_0|/|z|$ for all the deformations of the integration parth.

To prove that the order of integration can be interchanged, it is sufficient to check that the double integral converges absolutely and apply Fubini's theorem. Indeed, using the bound in \Eqref{eq_connection_B_bound}, we have
\begin{equation}
\int_0^{\infty e^{i\theta}} \! |\dd\zeta|\, e^{-\Re(z\zeta)}\, |B(\zeta)| \le K \int_0^\infty \dd |\zeta|\, e^{-\big(\Re(ze^{i\theta})-A\cos(\theta)\big)|\zeta|},
\end{equation}
and the last integral converges in $\Re\big( z e^{i\theta} \big)>A\cos(\theta)$.

%%%%%%%%%%%%%%%%%%%%%%%%%%%%%%%%%%%%%%%%%%%%%%%%%%%%%%%%%%%%%%%%%%%%%%%%%%%%%%%%%%%%%%%%%%%%%%%%%%%

\section{\texorpdfstring{Proof of \Propref{prop_resurgence_1} with $S \in \mathbb{C}\setminus\mathbb{R}^+$}{Proof of \Propref{prop_resurgence_1} with S in the complex plane minus the positive real axis}}
\label{sec_connection_proof_1'}

Here we will prove the following generalisation of \Propref{prop_resurgence_1}:
%Proof 1 generalised
\begin{prop}
\label{prop_resurgence_3}
Let $f$ be an analytic function in $\mathbb{C}\setminus(\mathbb{R}^\minus +A)$ and satisfy $|f(z)|\le |a_0|/|z|$ in $\Re(z)>A$ and $|f(z)|\le Ke^{s|z|}$ in $\Re(z)<A$\footnote{When $\Re(S)>0$, we expect that $f$ becomes exponentially enhanced somewhere in $\Re(z)<A$. Thus, imposing the bound $|f(z)|\le |a_0|/|z|$ on both half-planes, as we did in \Propref{prop_resurgence_1}, would be too restrictive.} (minus a neighbourhood of $A$ in both cases). Further assume that $\disc{f}$ satisfies \Eqref{eq_connection_disc_prop} with $S\in \mathbb{C}\setminus\mathbb{R}^+$ and the $\mathcal{O}(1/z)$ terms satisfy \Eqref{eq_connsection_bound_2}. Then
\begin{equation}
f(z) = \int_0^{\infty} \! \dd\zeta \, e^{-z\zeta}\, B(\zeta) \sim \sum_{n\ge 0} \frac{a_n}{z^{n+1}}\,, \quad \Re(z) > a\,,
\label{eq_connection_laplace_prop3}
\end{equation}
(the asymptotic expansion being uniformly valid) where $B$ is defined in \Eqref{eq_connection_borel_prop} and the coefficients $a_n$ satisfy \Eqref{eq_connection_an_prop}.
\end{prop}

The proof would go as follows. We consider the Borel transform $B$ in \Eqref{eq_connection_borel_prop}. Using the bound $|f(z)|\le Ke^{s|z|}$, valid in $\Re(z)<A$, we can write $B$ as in \Eqref{eq_connection_B_def2_proof1} and check that the function is analytic in $\Re(\zeta)>\Re(S)$.

The bounds in \Eqref{eq_connsection_bound_1} and \Eqref{eq_connsection_bound_3} are still valid, but \Eqref{eq_connsection_bound_2} might not due to the fact that $S$ is now complex. This is the reason we are forced to impose this bound in the assumptions of \Propref{prop_resurgence_3}. We find
\begin{equation}
B^{(n)}(0) = \frac{(-1)^{n+1}}{\pi} \frac{\Gamma(n+\alpha)}{(-S)^{n+\alpha}}\, b_0 \left[ 1 + \mathcal{O}\left( \frac{1}{n} \right) \right],
\end{equation}
which means that $B$ is also analytic in a disc of radius $|S|$ around the origin.

Finally, to prove \Eqref{eq_connection_laplace_prop3}, we repeat the same steps in \Appref{sec_connection_proof_1} (but only for $\theta=0$). The hypothesis that $|f(z)|\le |a_0|/|z|$ in $\Re(z)>A$ is used in the last line of \Eqref{eq_connection_B=f} to deform the path $\mathcal{C}_a(0)$ into a circle around $w=z$.

%\begin{figure}
%\centering
%\begin{tikzpicture}[scale=1]
%\fill[fill=gray!20!white] (1,0) arc(0:360:1);
%\fill[fill=gray!20!white] (0.75,-2) -- (0.75,2) arc(90:-90:2);

%\fill (0.75,0.661438) circle[radius=2pt] node[left]{$S$};

%\draw[dashed] (0,0) -- (2,1.76383);
%\draw[dashed] (0,0) -- (2,-1.76383);

%\draw[help lines,->] (-2,0) -- (2,0) coordinate (xaxis);
%\draw[help lines,->] (0,-2) -- (0,2) coordinate (yaxis);

%\path[draw,line width=0.8pt,decoration={markings, mark=at position 0.75 with {\arrow{>}}}, postaction=decorate] (0,0) -- (2.5,1);

%\draw[->] (1,0) arc (0:20.8014:1) node[midway,right]{$\theta$};

%\node[above] at (yaxis) {$\zeta$ plane};
%\end{tikzpicture}
%\caption{Region where $B$ is analytic in the proof of \Appref{sec_connection_proof_1'}. $S$ is a singularity of $B$ at the intersection of the half plane $\Re(\zeta)>S$ and a disc of radius $|S|$ around the origin.
%The dashed lines are the restrictions on $\theta$ for the directional Laplace transform.
%}
%\label{fig_connection_sketch_1'}
%\end{figure}

%%%%%%%%%%%%%%%%%%%%%%%%%%%%%%%%%%%%%%%%%%%%%%%%%%%%%%%%%%%%%%%%%%%%%%%%%%%%%%%%%%%%%%%%%%%%%%%%%%%

\section{Proof of \Propref{prop_resurgence_2}}
\label{sec_connection_proof_2}
%Proof 2
We define $f_\theta$ as the Laplace transform of $B$ along the path in \Figref{fig_connection_def_ftheta}, with $\theta\in [-\pi/2-\epsilon,\pi/2+\epsilon]$ (going around $S$ if necessary). Using the exponential bound on $B$, it is easy to check that $f_\theta$ is an analytic function in $\Re\big(z e^{i\theta}\big)>A$. Furthermore, the functions $f_\theta$ coincide in the intersection of the half-planes of analyticity. Therefore, concatenating the half-planes, we can analytically continue $f=f_0$ around the disc $D(0,A)$ (of radius $A$ and centre 0).

We consider the two directions $\theta_\minus = -\pi/2-\epsilon$ and $\theta_\plus = +\pi/2+\epsilon$. These directions define a pair of Laplace transforms whose difference can be written as
\begin{equation}
f_{\theta_\minus}(z) - f_{\theta_\plus}(z) = \int_\mathcal{C} \dd\zeta\, e^{-z\zeta}\, B(\zeta)\,,
\end{equation}
where $\mathcal{C} = \mathcal{C}_+ + \mathcal{C}_- + C_\delta$ is the contour in \Figref{fig_connection_disc_f}. This follows from a convenient deformation (if necessary) of the original paths that define $f_\theta$ and the fact that paths in opposite directions cancel each other.

We check that the integrals along $\mathcal{C}_\pm$ are $\mathcal{O}\big(e^{S'z}\big)$ for $z \rightarrow -\infty$, where $S'=S-\delta$. Indeed, given the parametrisation $\zeta = S' + xe^{i\theta_\pm}$, with $x\in \mathbb{R}^+$,
\begin{equation}
\begin{aligned}
\left| \int_{\mathcal{C}_\pm}\!  \dd\zeta\, e^{-z\zeta}\, B(\zeta)  \right| &\le e^{\Re(S')z} \int_0^\infty \dd x\, e^{-zx\cos(\theta_\pm)}\, \left|B\left(S'+xe^{i\theta_\pm}\right)\right|\\
&\le K e^{\Re(S')z} \int_0^\infty \dd x\, e^{-x(z\cos(\theta_\pm) - A)} = \mathcal{O}\big(e^{S'z}\big)\,.
\label{eq_connection_bound_C+-}
\end{aligned}
\end{equation}
Here we have used the bound $|B(\zeta)| \le Ke^{A|\zeta|}$ and verified that $z\cos(\theta_\pm) - A > 0$ for large enough $-z$. The fact that the integral is convergent also proves that $f_{\theta_\minus}(z) - f_{\theta_\plus}(z)$ is well defined for large enough $-z$.

$f_{\theta_\minus}(z) - f_{\theta_\plus}(z)$ is the difference between the two possible analytical continuations around $D(0,A)$. As we saw on \Exref{example_connection_resurgence}, this corresponds to
\begin{equation}
\disc{f}(z) = f_{\theta_\minus}(z) - f_{\theta_\plus}(z)\,, \quad z<0 \text{ and large enough.}
\end{equation}
This is an exact result at this stage, rather than an approximation.

From \Eqref{eq_connection_bound_C+-}, we see that the leading behaviour in \Eqref{eq_connection_disc_prop_2} can only come from the integral around $C_\delta$. Choosing $\delta$ small enough, we might use \Eqref{eq_connection_B_prop_2} to obtain
\begin{equation}
\begin{aligned}
\int_{C_\delta} \! \dd\zeta\, e^{-z\zeta}\, B(\zeta) &= \int_{C_\delta} \! \dd\zeta\, e^{-z\zeta}\left( \frac{-b_0}{\pi} \frac{\Gamma(\alpha)}{(\zeta-S)^\alpha} \bigg[ 1 + R(\zeta) \bigg] \right)\\
&=
\begin{multlined}[t][0.688\textwidth]
2i\, b_0\, e^{-Sz}(-z)^{\alpha-1} \Bigg[ 1 + \frac{\sin(\pi\alpha)\Gamma(\alpha)\Gamma(1-\alpha,-\delta z)}{\pi}\\
- \frac{1}{(-z)^{\alpha-1}}\frac{\Gamma(\alpha)}{2\pi i}\int_{C_{\delta,0}}\! \dd\zeta\, e^{-z\zeta}\, \frac{R(\zeta+S)}{\zeta^\alpha}  \Bigg],
\end{multlined}
\end{aligned}
\end{equation}
where $|R(\zeta + S)| \le K|\zeta|$, $\Gamma(s,x)$ is the upper incomplete gamma function and $C_{\delta,0}$ is a circle of radius $\delta$ around 0.
\begin{itemize}
\item Using the asymptotic behaviour $\Gamma(s,x) \sim x^{s-1}e^{-x}$ for large $x$, the first term is bounded by
\begin{equation}
\left| \frac{\sin(\pi\alpha)\Gamma(\alpha)\Gamma(1-\alpha,-\delta z)}{\pi} \right| \le K_1 \frac{e^{\delta z}}{(-\delta z)^\alpha} = \mathcal{O}\left( \frac{1}{z} \right).
\end{equation}

\item Using $|R(\zeta + S)| \le K|\zeta|$, the second term is bounded by
\begin{equation}
\left| \frac{1}{(-z)^{\alpha-1}}\frac{\Gamma(\alpha)}{2\pi i}\int_{C_{\delta,0}}\! \dd\zeta\, e^{-z\zeta}\, \frac{R(\zeta+S)}{\zeta^\alpha} \right| = \mathcal{O}\left( \frac{1}{z} \right).
\end{equation}
\end{itemize}

From the above results, \Eqref{eq_connection_disc_prop_2} is realised. Notice that the first bound can be improved to an arbitrary power of $1/z$. Actually only the second bound gives the error in \Eqref{eq_connection_disc_prop_2}.

%%%%%%%%%%%%%%%%%%%%%%%%%%%%%%%%%%%%%%%%%%%%%%%%%%%%%%%%%%%%%%%%%%%%%%%%%%%%%%%%%%%%%%%%%%%%%%%%%%%

\section{A path integral with no resurgent connection}
\label{sec_connection_resurgence_fails}
We consider the self-energy $\Sigma$ in the $O(N)$ non-linear sigma model. To next-to-leading order in $1/N$, it is given by
\begin{equation}
\Sigma\big(p^2\big) = \frac{1}{\pi N} \int_{\mathbb{R}^2} \dd^2 k\, \frac{\sqrt{k^2(k^2+4m^2)}}{\log\left[ \frac{\sqrt{k^2 + 4m^2}+\sqrt{k^2}}{\sqrt{k^2 + 4m^2}-\sqrt{k^2}} \right]} \frac{1}{(p+k)^2 + m^2}\,,
\label{eq_connection_sigma_Sigma}
\end{equation}
where $m^2 = \mu^2 e^{-1/g(\mu)}$ is the dinamically generated mass of the $\sigma$ particle and $g(\mu)$ is the coupling of the model at the scale $\mu$. (We follow the same notation as in \cite{BenekeBraunKivel}).

For convenience, we define the variable $z=1/g(p)$ and the dimensionless function
\begin{equation}
E(z) = \frac{N \Sigma_R\big(m^2 e^z\big)}{m^2 e^z}\,,
\end{equation}
where $\Sigma_R$ is the renormalised self-energy $\Sigma$, obtained after two zero-momentum subtractions. An asymptotic expansion for $E(z)$ is given by (see \cite[Eq.~17]{BenekeBraunKivel})
\begin{equation}
E(z) \sim -\log(z) + c + \widetilde{E}(z) = -\log z + c + \sum_{n\ge 0} \frac{\sigma_n}{z^{n+1}}\,,\quad z\rightarrow +\infty\,,
\label{eq_connection_sigma_E_asymptotic}
\end{equation}
where $c = 1.887537\dots$ and
\begin{equation}
\sigma_n =
\begin{cases}
-2 & \text{if } n=0\\
n! \big\{ [1+(-1)^n]\zeta(n+1) - 2 \big\} & \text{if } n\ge 1\,.
\end{cases}
\end{equation}
($\zeta$ is the Riemann $\zeta$-function). We can see the explicit factorial divergence in the coefficients $\sigma_n$ of the asymptotic expansion. The symbol $\widetilde{E}(z)$ contains only the power-like part of the asymptotic expansion of $E(z)$.

The Borel transform of $\widetilde{E}$ in \Eqref{eq_connection_sigma_E_asymptotic} is given by
\begin{equation}
\widehat{E}(t) =  \sum_{n\ge 0} \frac{\sigma_n t^n}{n!} = \frac{1}{t-1} - \psi(1+t) - \psi(2-t) - 2\gamma\,,
\label{eq_median_BT}
\end{equation}
where $\psi$ is the digamma function and $\gamma$ is the Euler constant. $\widehat{E}(t)$ has simple poles at $t = k \in \mathbb{Z}\setminus \{0\}$ and is analytic elsewhere. The residue of the poles along the positive real axis are given by
\begin{equation}
r_k = \textrm{Res}\left( \widehat{E}(t),\, t=k \right) =
\begin{cases}
+1 & \text{if $k=1$}\\
-1 & \text{if $k\ge 2$.}
\end{cases}
\end{equation}

Because we expect that $E(z)$ is real for $z>0$ ($g>0$), we consider the median resummed series
\begin{equation}
E^\textrm{mr}_0(z) = -\log(z) + c + \int_0^{\infty e^{i\theta}} \! \dd t\, e^{-zt}\, \widehat{E}(t) \pm i\pi \sum_{k\ge 1} r_k e^{-kz}\,,
\label{eq_connection_sigma_MR}
\end{equation}
where the minus sign is chosen when $\theta \in (0,+\pi)$ and the plus sign, when $\theta\in (-\pi,0)$.

Later we will argue that $E^\textrm{mr}_0$ does not coincide with $E$, but still we want to understand the properties of $E^\textrm{mr}_0$, because the singularties in the Borel plane of median resummed series determine the discontinuity of the function (\Propref{prop_resurgence_2}).

Our first observation is that, from $\psi(t) \sim \log(t)$, valid for large $|t|$, we resolve that $\big|\widehat{E}(t)\big| \sim 2\log|t|$. This implies that the Laplace transform in \Eqref{eq_connection_sigma_MR} defines an analytic function in the half-planes $\Re\big( ze^{i\theta}\big) >0$ (as a consequence of part \ref{it_connection_nevanlinna(b)} of Nevanlinna's theorem).

By choosing directions $\theta \in (-\pi/2,+\pi/2)$, $\theta\neq 0$, and concatenating the half-planes of analyticity, $E^\textrm{mr}_0(z)$ becomes an analytic function in $\mathbb{C}\setminus \mathbb{R}^-$. When changing the direction from below to above the positive real axis of the Borel plane, the two resulting functions coincide in the intersection of the half-planes, thanks to the exponential terms arising from median resummation. Therefore, they provide an analytic continuation of one another (even if the singularities along the positive real axis would normally prevent that).

From the discussion of \Secref{sec_connection_median}, we concluded that we might use the result of \Propref{prop_resurgence_2} for median resummed series. That is, we may obtain the asymptotic behaviour of $\disc{E^\textrm{mr}}_0(z)$ for large $-z$ from the singularities in the Borel transform.

However, here we face a problem that we did not realise in any of our previous examples. The asymptotic behaviour of the discontinuity is fixed by the singularity with the largest real part  in the Borel plane, but in this example there is no such singularity. We have an infinite amount of singularities along the positive axis, each with a larger real part than the previous.

One way to proceed is to compute the contribution to the discontinuity from all singularities along the positive real axis, sum the series in $\Re(z)>0$, analytically continue the result to $\Re(z)<0$ and only then extract the leading behaviour for large $-z$.

The contribution to the discontinuity from all positive singularities is given by
\begin{equation}
2\pi i \sum_{k\ge 1} r_k e^{-kz} = 2\pi i \frac{e^{-z}\big( 1 - 2e^{-z} \big)}{1-e^{-z}}\,.
\label{eq_connection_sigma_residue_sum}
\end{equation}
While originally the series on the left only converges in $\Re(z)>0$, the closed form on the right provides an analytic continuation to $\Re(z)<0$. The leading behaviour of \Eqref{eq_connection_sigma_residue_sum} in $\Re(z)<0$ yields
\begin{equation}
\disc{E^\textrm{mr}_0}(z) \sim (2\pi i) 2 e^{-z}\,, \quad z\rightarrow -\infty\,.
\label{eq_connection_sigma_disc_MR}
\end{equation}
Effectively, it is as if the singularity with the largest real part were a simple pole at $t=1$ with residue 2. It is easy to check that the discontinuity from the explicit logarithm in \Eqref{eq_connection_sigma_MR} and the contributions from negative singularities yield sub-leading corrections to \Eqref{eq_connection_sigma_disc_MR}.

%This computation already shows that $E^\textrm{mr}_0$ does not feature a connection between the large order behaviour of the coefficients $\sigma_n$ and the asymptotic expansion of $\disc{E^\textrm{mr}_0}$. As we said in the paragraph before \Propref{prop_resurgence_1}, the large order behaviour of the $\sigma_n$ determines the closest singularities to the origin of the Borel plane, in this case, at $t=\pm 1$. On the other hand, the asymptotic behaviour of $\disc{E^\textrm{mr}_0}$ is fixed from the singularity with the largest real part, which in this example comes from the combination of all singularities along the positive real axis.

While \Eqref{eq_connection_sigma_disc_MR} correctly encodes the discontinuity of $E_0^\textrm{mr}$, this result is invalid for the exact function $E$. We will prove that there are additional exponential corrections to \Eqref{eq_connection_sigma_disc_MR} which are not captured by median resummation. These exponential corrections will contribute to $\disc{E}$ on top of \Eqref{eq_connection_sigma_disc_MR}. Therefore, it is impossible that $E$ satisfies a resurgent connection.

The full \guillemotleft Borel representation\guillemotright{} of $E$ is given by (see \cite[ Eq.~14]{BenekeBraunKivel}):
\begin{equation}
E(z) = \int_0^{\infty e^{i\theta}}\! \dd t \sum_{n \ge 0} (-1)^n e^{-nz} \left( e^{-zt} \Big[ z\, F_n(t) + G_n(t) \Big] - H_n(t) \right),
\label{eq_connection_sigma_beneke}
\end{equation}
where we have introduced a direction $\theta$ in the integral in order to incorporate our framework to the discussion.

The functions $F_n$, $G_n$ and $H_n$ can be found in \cite[App.]{BenekeBraunKivel}. We quote their expressions for $n=0$:
\begin{align}
F_0(t) &= 1\,,\\
G_0(t) &= \frac{1}{t} + \frac{1}{t-1} - \psi(1+t) - \psi(2-t) - 2\gamma\,,\\
H_0(t) &= \frac{1}{t} + B_1(t)\,,
\end{align}
where $B_1$ is a function analytic in $\mathbb{C}\setminus (\mathbb{R}^\minus - 2)$.

The integral in \Eqref{eq_connection_sigma_beneke} is well-defined for $\theta=0$ despite the poles present in $G_n$ and $H_n$ along the positive real axis. The poles completely cancel each other in the sum over $n$. Namely, the cancellation of the pole at $t=t_0$ occurs between $G_n(t)$ and $H_{n+t_0}(t)$. This is the process of renormalon cancellation.

It can be checked that the pole at the position $t=k$ coming from $H_k$ provides the exponential correction $r_k e^{-kz}$ in \Eqref{eq_connection_sigma_MR}. In the same way, the sign changes with the direction $\theta$ chosen in \Eqref{eq_connection_sigma_beneke}. Thus, this representation already incorporates the exponential corrections arising from median resummation. Still, \Eqref{eq_connection_sigma_beneke} contains additional corrections. Let us make this explicit.

Considering only $n=0$ in \Eqref{eq_connection_sigma_beneke}, we have
\begin{multline}
\int_0^{\infty e^{i\theta}}\! \dd t \left( e^{-zt} \Big[ z\, F_0(t) + G_0(t) \Big] - H_0(t) \right) = 1 + \int_0^{\infty e^{i\theta}} \! \dd t\, e^{-zt}\, \widehat{E}(t)\\
- \int_0^{\infty e^{i\theta}} \! \dd t \left( H_0(t) - \frac{e^{-zt}}{t} \right).
\end{multline}
The last integral yields\footnote{From this computation we come to the conclusion that the pole at $t=0$ in $G_0(t)$ has nothing to do with renormalisation, as claimed in \cite{BenekeBraunKivel}. Instead, the pole at $t=0$ encodes the $\log$ term appearing in the asymptotic expansion of $E(z)$ (see \Eqref{eq_connection_sigma_E_asymptotic}). Note that applying \Eqref{eq_connection_borel} with $f(z) = \log(z)$ yields a pole at 0.}
\begin{equation}
- \int_0^{\infty e^{i\theta}} \! \dd t \left( H_0(t) - \frac{e^{-zt}}{t} \right) = -\log(z) + c - 1\,.
\end{equation}
Thus, we verify that the term $n=0$ is equal to \Eqref{eq_connection_sigma_MR} up to the ambiguous exponential terms arising from median resummation. (We are missing them because they come from $H_n$, with $n\ge 1$).

To finish the discussion, it is a simple verification that the poles in $G_1$ contribute to the discontinuity of $E(z)$. A similar computation to that in \Eqref{eq_connection_sigma_residue_sum} and \Eqref{eq_connection_sigma_disc_MR} yields
\begin{equation}
\disc{E^\textrm{mr}}_1(z) \sim (2\pi i)2e^{-z}\,, \quad z\rightarrow -\infty\,,
\label{eq_connection_sigma_disc_MR_2}
\end{equation}
where $E^\textrm{mr}_1$ is the term $n=1$ in \Eqref{eq_connection_sigma_beneke} with the necessary exponential corrections to cancel the imaginary ambiguities:
\begin{equation}
E^\textrm{mr}_1(z) = -e^{-z} \int_0^{\infty e^{i\theta}}\! \dd t \left( e^{-zt} \Big[ z\, F_1(t) + G_1(t) \Big] - H_1(t) \right) \pm i\pi\, \text{(exponentials).}
\end{equation}

This clarifies that, to obtain the real asymptotic behaviour for $\disc{E}$, the result in \Eqref{eq_connection_sigma_disc_MR} has to be corrected by adding the contribution in \Eqref{eq_connection_sigma_disc_MR_2} and, actually, by adding all contributions from the terms $n\ge 1$.

\bibliographystyle{utphys}
\bibliography{references}

\end{document}